\documentclass[%
 reprint,
 amsmath,amssymb,
 aps,
pra,
]{revtex4-2}

\usepackage{comment}
\usepackage[normalem]{ulem}
\usepackage{graphicx}
\usepackage{dcolumn}
\usepackage{bm}
\usepackage{hyperref}
\usepackage{soul}
\usepackage[dvipsnames]{xcolor}
\setstcolor{red}

\begin{document}

\title{Coarse-Graining via Lumping: Exact Calculations and Fundamental Limitations}

\author{Gianluca Teza}
\email{teza@pks.mpg.de}
\affiliation{Max Planck Institute for the Physics of Complex Systems, N\"othnitzer Str.~38, 01187 Dresden, Germany}
\author{Attilio L. Stella}
\email{stella@pd.infn.it}
\affiliation{Department of Physics and Astronomy, University of Padova, and INFN, Sezione di Padova, Via Marzolo 8, I-35131 Padova, Italy}
\author{Trevor GrandPre}
\email{trevorg@wustl.edu}
\affiliation{Department of Physics, Washington University in St. Louis, St. Louis, MO 63130 USA}
\affiliation{National Institute for Theory and Mathematics in Biology, Northwestern University and The University of Chicago, Chicago, IL, USA.}

\date{\today}

\begin{abstract}
Detecting broken time-reversibility at micro- and nanoscale is often difficult when experiments offer limited state resolution. We introduce a lumping method that builds an effective semi-Markov model able to reproduce exactly the full entropy-production statistics of the microscopic dynamics. The mean entropy production stays accurate even when hidden current-carrying cycles are merged, though higher-order information can be unavoidably lost. In these cases, we capture violations of fluctuation theorems consistent with experiments, opening a path to novel inference strategies out of equilibrium.
\end{abstract}

\maketitle

Stochastic thermodynamics provides a unified framework for describing small systems driven out of equilibrium \cite{seifert2012stochastic}. In this framework, fluctuating quantities such as currents, work, and entropy production obey fundamental relations, including fluctuation–dissipation and fluctuation–response relations~\cite{speck2006restoring, prost2009generalized, baiesi2009fluctuations, seifert2010fluctuation, chetrite2011two, baiesi2013update, maes2014second, kwon2025fluctuation, dechant2020fluctuation}, fluctuation theorems~\cite{lebowitz1999gallavotti, jarzynski2006rare, jarzynski1997nonequilibrium, searles1999fluctuation, kurchan1998fluctuation, crooks1999entropy, seifert2005entropy}, and thermodynamic~\cite{horowitz2020thermodynamic, gingrich2017fundamental, barato2015thermodynamic, gingrich2016dissipation, polettini2016tightening, dechant2025inverse, raghu2025thermodynamic} as well as kinetic uncertainty relations~\cite{di2018kinetic, van2022unified, palmqvist2025kinetic, hiura2021kinetic, yan2019kinetic}.
These results depend critically on an accurate characterization of fluctuations of relevant observables (e.g. density-like and current-like observables)~\cite{seifert2025universal}
When experimental or observational constraints restrict the accessible state space, a fundamental question arises: to what extent does a coarse-grained description preserve the structure of stochastic thermodynamics, and under what conditions can it remain exact, retaining all information of the original dynamics?

Coarse-graining arises when only part of a system’s microscopic dynamics can be experimentally resolved or when groups of states are indistinguishable. Although many approaches provide approximate coarse-grained descriptions or rely on limiting assumptions~\cite{martinez2019inferring, van2023time, harunari2022learn, ehrich2021tightest, ghosal2022inferring, pietzonka2024thermodynamic, baiesi2024effective, skinner2021estimating, blom2024milestoning,diterlizzi2025forcefree}, whether a coarsened Markovian description can ever be exact in general remains unresolved. Nevertheless, it is valuable to develop principled criteria for when particular classes of coarse graining are expected to be valid.

In Markov processes on discrete state spaces, a paradigmatic and experimentally supported setting for stochastic thermodynamics \cite{pekola2015towards,ciliberto2017,alemany2015,gnesotto2018}, common coarse-graining operations include decimation, which removes unobserved states while preserving transitions among the remaining ones, and lumping, which merges microstates into effective states. Both operations have clear biological counterparts: decimation naturally applies when some microscopic states or transitions are experimentally inaccessible or latent, as in single-molecule measurements or intracellular processes with hidden degrees of freedom \cite{mallick2007nonequilibrium,teza2020exact}. Lumping, by contrast, is appropriate when multiple microscopic states are functionally indistinguishable and collectively manifest as a single effective state, for example due to limited experimental resolution, observational equivalence, or the presence of multiple time scales that motivate grouping fast and slow processes \cite{dieball2025perspective,geiger2014lumpings,hoffmann2009bounding,igoshin2025uncovering,esposito2012stochastic,rahav2007fluctuation,amann2010communications,nicolis2011transformation,hartich2023violation, tabanera2025purely}. Semi-Markov descriptions~\cite{hughes1995random,esposito2008continuous,andrieux2008thefluctuation} arise naturally under decimation~\cite{teza2020exact}, yielding an exact coarse graining that preserves the full statistics of entropy production; whether this result extends to the broader and more mathematically challenging class of coarse graining associated with lumping, however, remains unclear.

Full knowledge of the microscopic Markov dynamics -- namely the stationary probabilities and transition rates among all states -- enables the exact computation of the steady-state entropy production rate (EPR) as an average of the relative entropies between all trajectories
and their time-reverses~\cite{schnakenberg1976network, van2015ensemble, mallick2009some, peliti2021stochastic, ziener2015entropy, seifert2005entropy, gaspard2004time, maes2003time, parrondo2009entropy}.
In this settings, large deviation theory provides a framework to also access fluctuations and higher order statistics \cite{varadhan2010large,touchette2011basic}.
Coarse-graining introduces memory effects and hidden currents in the dynamics, which make it difficult to characterize the out-of-equilibrium properties of the system, and to test whether fluctuation relations like the Gallavotti-Cohen symmetry remain valid~\cite{lebowitz1999gallavotti}.

Here, we introduce a coarse-graining strategy via lumping to showcase how microscopic entropy production distributions can be retained \textit{exactly} through an appropriate semi-Markovian description, as previously proved for decimation~\cite{teza2020exact}.
We also show how collapsing loops carrying net amount of currents can make the lumped description no longer exact for the full EPR distribution.
Nevertheless, the average value of the EPR--the key observable to discriminate between in- and out-out-of-equilibrium systems--can be retained exactly in such cases.
Lastly, we show how this lumping description captures violations of the thermodynamic uncertainty relation and the breaking of Gallavotti-Cohen symmetries~\cite{lebowitz1999gallavotti} observed in experiments at the micro and nano scale~\cite{mehl2012role,battle2016broken,diterlizzi2024variance}, opening the way to novel inference strategies for systems outside equilibrium.

Let us start with the paradigmatic example of a driven particle on a ring system, that can be effectively described with a memory-less (Markovian) dynamics on a periodic linear network of $N$ state configurations. The probability $P_i(t,S)$ of observing the system on the $i$-th configuration at a time $t$ with a total accumulated entropy $S$ is regulated by the master equation (ME) \cite{teza2020exact}:
\begin{equation}\label{eq:ME_generic_network}
    [\partial_t + \sum_{j\neq i} W_{ji} ] P_i(t,S)= \sum_{j\neq i} W_{ij} P_j\left(t,S-\log \frac{W_{ij}}{W_{ji}}\right) 
\end{equation}
where $W_{ij}$ is the transition rate from the $j$-th to the $i$-th state and summing over all $S$ one recovers a standard ME for diffusion.
Each transition $i\to j$ provides a contribution $\log W_{ij}/W_{ji}$ to the overall produced entropy $S$ \cite{lebowitz1999gallavotti,seifert2005entropy}, such that for every trajectory in the steady state we have that $P(S,t)\equiv \sum_i P_i(S,t)$ is consistent with a large deviation principle  for the entropy production rate (EPR) $\sigma=S/t$ \cite{touchette2009large}.
This means that for $t\to\infty$ the probability $P(S,t)$ concentrates around the value $S=\left< \sigma \right> t$ where 
\begin{equation}
    \left< \sigma \right> =\sum_{i,j\neq i}^N W_{ij} P^*_j \log \frac{W_{ij}P^*_j}{W_{ji}P^*_i} .
\end{equation}
is the average scaled EPR and $P^*_i$ are the steady state probabilities solving Eq. \ref{eq:ME_generic_network} for $t\to\infty$. \cite{schnakenberg1976network,van2015ensemble,mallick2009some,peliti2021stochastic,ziener2015entropy}.

If the configurations can be ascribed, e.g., to the position of a particle in a lattice of spacing $L$, we can assume that the system can be driven outside of equilibrium through some uniform drive $f$ pushing the particle along the network with jump rates $r$ and $l$ between nearest neighboring right and left sites.
The rates satisfy a local DB condition $r/l=e^{\beta L f}$ \cite{maes2021local,teza2020exact,teza2020rate}, where $\beta=1/k_B T$ is the thermal bath temperature and $k_B$ is the Boltzmann constant. 
The ME (Eq. \ref{eq:ME_generic_network}) in this scenario reduces to
\begin{eqnarray} \nonumber
    [\partial_t +r+l]P_i(t,S)&=r P_{i-1}\left(t,S-\log \frac{r}{l}\right) \\
    &+ l P_{i+1}\left(t,S-\log \frac{l}{r}\right)
\end{eqnarray}
which allows us to access all the EPR statistics through the evaluation of the corresponding generating function $G(\lambda,t)=\sum_{i,S} e^{\lambda S} P_i(t,S)$.
Indeed, the large deviation principle implies that in the long-time limit the leading part of the cumulants generator $\log G(\lambda,t)$ is extensive in time \cite{stella2023anomalous,stella2023universal,teza2025universal}, meaning that we can evaluate the scaled cumulant generating function (SCGF) from
\begin{equation}\label{eq:SCGF_ring_true}
    \varepsilon(\lambda)=\lim_{t\to\infty} \frac{\log G(\lambda,t)}{t}=re^{\lambda \log r/l} +le^{\lambda \log l/r} -r-l
\end{equation}
giving access to all the scaled cumulants upon differentiation with respect to $\lambda$ at $\lambda=0$.

\begin{figure}
    \centering
    \includegraphics[width=0.95\linewidth]{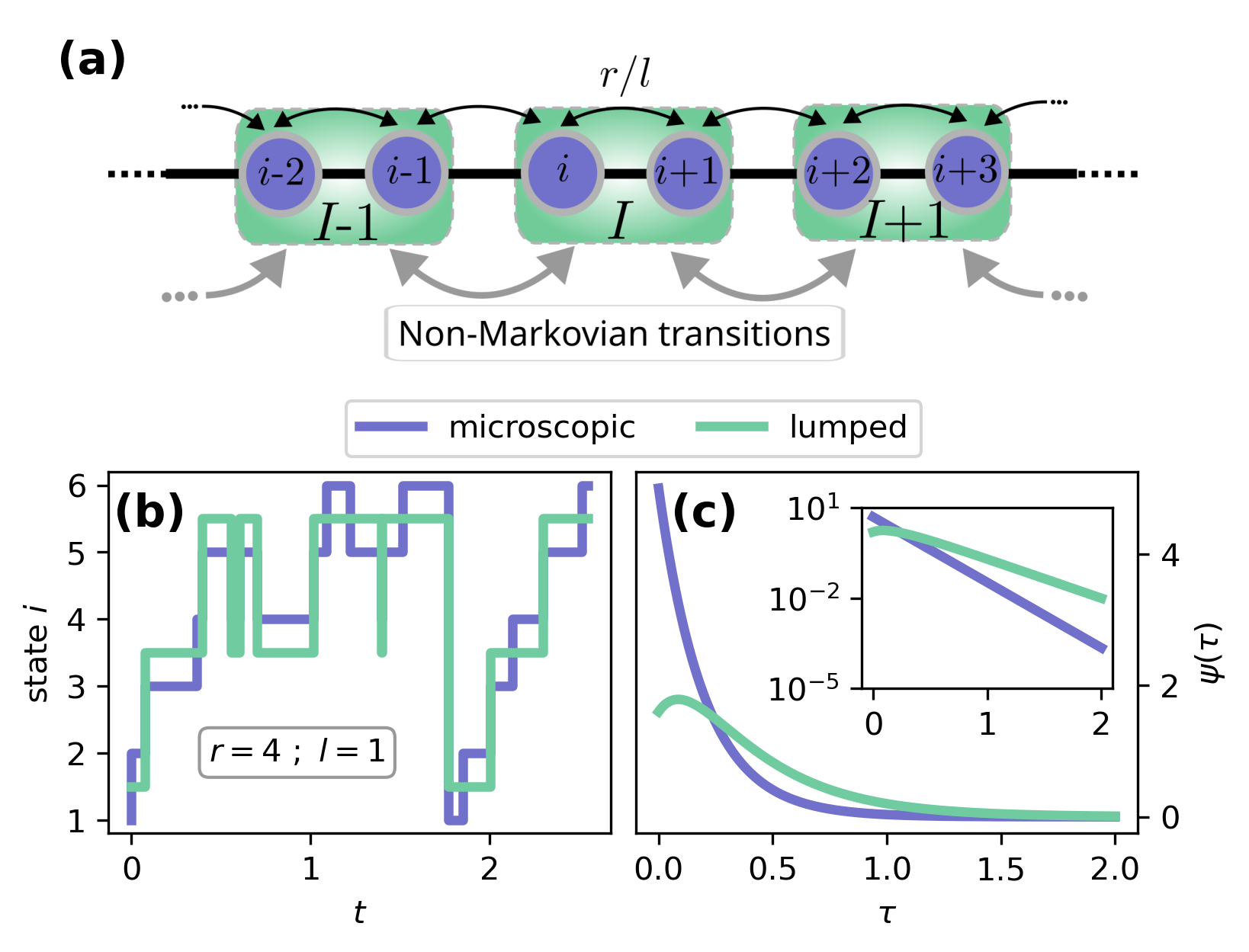}
    \caption{Lumping pairs of states on a one-dimensional ring introduces memory.
    (a) A one-dimensional Markov ring with right and left hopping rates $r$ and $l$, respectively. Adjacent sites are lumped in pairs, producing a non-Markovian coarse-grained system where memory arises from unresolved microscopic transitions. 
    (b) Example of a coarse-grained trajectory (green) obtained by lumping a microscopic trajectory (blue) on a six-state ring with $r=4$ and $l=1$. 
    (c) Waiting-time distributions of the microscopic (blue) and lumped (green) systems. The lumped distribution $\Psi(t)$ (Eq.~\ref{eq:Psi_uniform_ring}) deviates from the exponential form of the Markov process, reflecting memory effects; the inset shows the same curves on a log-linear scale.}
    \label{fig:lumping_sketch}
\end{figure}

We now coarse-grain the system by lumping adjacent sites in pairs: if $i$ is even, sites $i$ and $i+1$ are merged into a single coarse-grained state $I$ (see Fig.~\ref{fig:lumping_sketch}a). Capital indices denote lumped states. The same procedure can be extended to merge any number of sites. Experimentally, this type of coarse-graining corresponds to an inability to distinguish between microscopic configurations within each lumped state. This situation differs fundamentally from decimation, where specific microscopic transitions are unobserved~\cite{teza2020exact}, leading to a distinct coarse-grained dynamics (see Fig.~\ref{fig:lumping_sketch}b). Mathematically, this difference arises because lumping cannot be implemented by direct algebraic elimination in Eqs.~\ref{eq:ME_generic_network}. The resulting dynamics cannot be expressed as a closed system involving only the variables $P_I = P_i + P_{i+1}$, and thus requires a different construction, outlined below.

In the microscopic Markov process, the waiting-time (i.e., sojourn-time) distributions are exponential, $\psi_i(t) = A_i^{-1} e^{-t/A_i}$, with characteristic times $A_i = (\sum_{j \neq i} W_{ji})^{-1}$. In the lumped description, however, the inability to resolve transitions among the constituent microstates introduces memory effects, leading to non-exponential waiting-time distributions (see Fig.~\ref{fig:lumping_sketch}c). The situation is further complicated because the possible trajectories depend on the microstate through which the system enters or exits the lumped state.

For the lumping scheme described above, analyzing the waiting times between entry and exit of the coarse-grained states allows one to construct effective dynamical equations that reproduce exactly the microscopic entropy-production distribution at stationarity. Computing the waiting-time distribution for a given pair of entry and exit microstates is straightforward, but incorporating the correlations between successive entries and exits along a coarse-grained trajectory is far more challenging. Since our focus is on the steady-state entropy production rate, we avoid this complication by formulating equations of motion with effective waiting-time distributions and transition rates that capture, in a cumulative sense, the long-time effects of the underlying microscopic transitions.

Introducing the Laplace time transform $\tilde{f}(u)=\int_{0^-}^{+\infty}dt\ e^{-ut} f(t)$, we can express a weighted waiting time distribution $\tilde{\Psi}(u)$ for all possible entrances and exits from a given lumped state.
Indicating with $\pi_r=l/(r+l)$ and $\pi_l=r/(r+l)$ the probabilities of entering the lumped state from the right and left, respectively, one finds (see the Supplemental Material for a full derivation and an analytic expression of $\Psi(t)$ \cite{SM}):
\begin{eqnarray}\label{eq:Psi_uniform_ring}
    \tilde{\Psi}(u)=\frac{\tilde{B}(u)\tilde{\psi}^2(u)}{(r+l)^2}\left[\pi_l r^2+\pi_r l^2+\frac{2rl}{\tilde{\psi}(u)}\right]
\end{eqnarray}
where $\tilde{\psi}(u)=(r+l)/(r+l+u)$ is the Laplace transform of the exponential waiting time distribution of the original model and $\tilde{B}(u)=\frac{(r+l)^2}{(r+l)^2-rl\tilde{\psi}^2(u)}$ is a term accounting for the probability of performing an indefinite number of jumps between the microstates composing the lumped state \cite{SM}.
We underline how terms appearing in the square bracket on the r.h.s. of Eq. \ref{eq:Psi_uniform_ring} are each linked to a specific transition: $r^2$ ($l^2$) accounts for a trajectory that has entered from the left (right) and exited to the right (left) of the lumped state, while the remaining term accounts for trajectories that entered and exited on the same side. 

In Laplace space, $\tilde{\Psi}(u)$ is directly connected to the memory kernel $\tilde{W}(u)=u\tilde{\Psi}(u)/(1-\tilde{\Psi}(u))$ regulating the dynamics of a generalized Master Equation (ME) \cite{montroll1965random,klafter1980derivation} on the lumped lattice.
Keeping into account the entropic contribution of each trajectory, the probability of having accumulated a total entropy $S$ while being on the state $I$ at a given $u$ evolves according to:
\begin{eqnarray}\label{eq:1D_lumped_GME} \nonumber
    \tilde{P}_I(u,S)-\tilde{c}(u)&= \frac{\tilde{B}(u)}{(r+l)(r+l+u)^2}[r^3 \tilde{P}_{I-1}(u,S-2\log \frac{r}{l}) \\ \nonumber
    &+l^3 \tilde{P}_{I+1}(u,S-2\log \frac{l}{r}) \\ \nonumber
    &+rl(r+l+u) \tilde{P}_{I-1}(u,S-\log \frac{r}{l}) \\ 
    &+rl(r+l+u) \tilde{P}_{I+1}(u,S-\log \frac{l}{r})]
\end{eqnarray}
where the term $\tilde{c}(u)=\frac{1-\tilde{\Psi}(u)}{u}P_I(t=0,S)$ retains information on the initial conditions and consequently won't play a role in determining the steady-state dynamics~\cite{diterlizzi2020explicit}.
Upon summing over $S$ and reverse-Laplace transforming Eq. \ref{eq:Psi_uniform_ring} one recovers a standard generalized ME for the dynamics (see Supplemental Material (SM)\cite{SM}).

Because $\tilde{\Psi}$ is constructed to resolve all microscopic entry–exit
trajectories through a lumped state, and the generalized master
equation~\ref{eq:1D_lumped_GME} assigns the corresponding entropy production to
each transition, the resulting dynamics reproduces exactly the total dissipation
accumulated along long coarse-grained trajectories. This will enable us to account for the correct EPR in the lumped description.
To do so, we inverse-Laplace transform Eq. \ref{eq:1D_lumped_GME} and obtain a linear second order time differential equation for the EPR generator $G'(\lambda,t)=\sum_{S,I} e^{\lambda S} P_I(t,S)$ of the lumped system \cite{SM}.
In the steady state, the large deviation principle ensures that $G'(\lambda,t)\sim e^{t\varepsilon'(\lambda)}$, which allows us to obtain the SCGF of the lumped system $\varepsilon'(\lambda)$ directly from the characteristic equation of the homogeneous differential equation of the EPR generator.
This yields:
\begin{eqnarray} \nonumber
    &\varepsilon'(\lambda)^2 + 2\varepsilon'(\lambda) \left[r+l-\frac{rl}{r+l} \cosh\left(\lambda \log \frac{r}{l}\right)\right] +(r+l)^2 \\
    &-rl\left(1+2\cosh\left(\lambda \frac{r}{l}\right)\right)-\frac{r^3e^{2\lambda \log \frac{r}{l}}+l^3e^{2\lambda \log \frac{l}{r}}}{r+l}=0 .
\end{eqnarray}

 A straightforward evaluation of the dominant root shows that $\varepsilon'(\lambda)\equiv\varepsilon(\lambda)$, so the lumped dynamics described by Eq.~\ref{eq:1D_lumped_GME} reproduces exactly the entire steady-state EPR distribution of the microscopic system. This is our first
striking result: we have constructed a coarse grained semi-Markov description that preserves not only the mean EPR but the full large-deviation spectrum.
From a large-deviation perspective, lumping contracts the microscopic path measure onto a reduced trajectory space via a many-to-one mapping on trajectories. Because the scaled cumulant generating function is a nonlinear functional of
this measure, such contractions generically alter large-deviation spectra—nonlinear operations do not, in general, commute with many-to-one
mappings~\cite{dembo2009large,greven1994large}. Exact preservation is therefore only expected in special cases where the coarse-grained dynamics
retains sufficient information to represent the same additive functional, as realized here by our exact semi-Markov construction.

The memory effects stemming from the lumping procedure are encoded in high order time derivatives as well as gradient mixed terms, which consist in time derivatives of probabilities not centered on the $I$-th site.
If one is only interested in preserving the average EPR, it is possible to restore a Markovian description of the evolution by neglecting higher order derivatives and expanding the gradient terms around the $I$-th state \cite{teza2020thesis,SM}.
This, however, does not generally preserve the correct amplitude of fluctuations and higher cumulants, which require an exact accounting of the memory effects.

In the Appendix we operate lumping on another periodic chain presenting secondary loops (Fig. \ref{figAP:secondary_loops_lumping}).
If secondary loops are collapsed into a single macro-state placed on the primary loop, an exact lumping scheme can be nevertheless performed to properly account for the EPR contributions produced in the hidden loops.
This allows to preserve the full EPR statistics yielding results that coincide with those obtained by decimation \cite{teza2020exact}.

\begin{figure}
    \centering
    \includegraphics[width=1.00\linewidth]{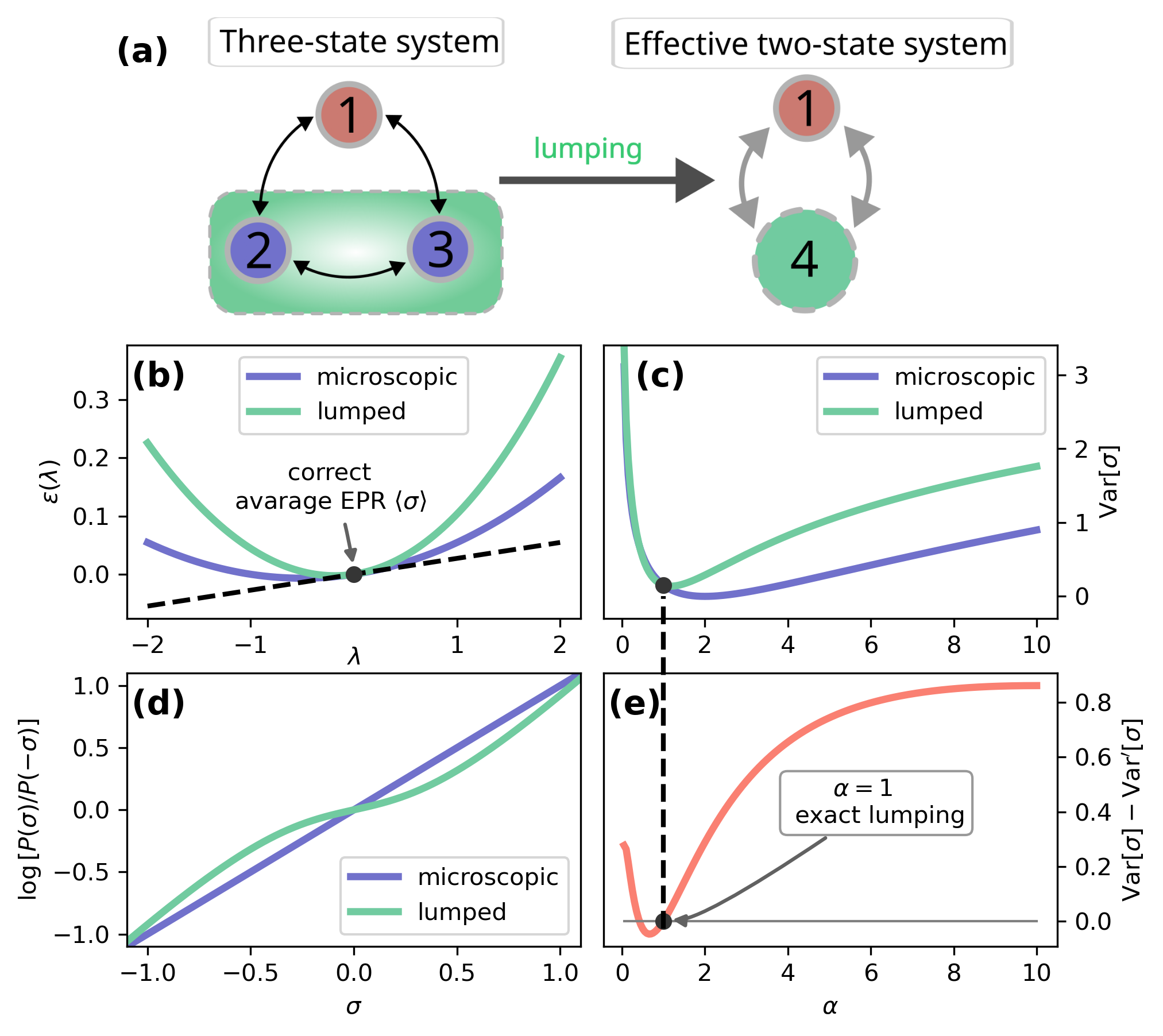}
    \caption{Lumping through loops and the breakdown of Gallavotti-Cohen Symmetry. 
(a) Three-state system coarse-grained into an effective two-state model. 
(b) Scaled cumulant generating functions (SCGFs) $\varepsilon(\lambda)$ of the microscopic (blue) and lumped (green) systems. The parameters are $W_{12}=3$, $W_{23}=2$, $W_{31}=1$ (counterclockwise) and $W_{13}=2$, $W_{32}=1$, $W_{21}=2$ (clockwise), placing the system out of equilibrium. Both SCGFs share the same slope at $\lambda=0$, $\partial_\lambda \varepsilon(\lambda)|_{\lambda=0} \equiv \langle\sigma\rangle$, showing that the coarse-grained model reproduces the exact mean EPR. 
(c) Broken Gallavotti--Cohen symmetry in the lumped system: the probability of observing negative entropy fluctuations deviates from the linear relation obeyed by the microscopic dynamics (blue). 
(d) Entropy-production fluctuations in the lumped system as a function of the parameter $\alpha$ in $W_{21}=\alpha\, W_{31}W_{12}/W_{13}$. 
(e) Difference in variances, $\mathrm{Var}[\sigma]-\mathrm{Var}'[\sigma]$, between the microscopic and lumped descriptions. 
At $\alpha=1$, the lumping becomes exact ($\varepsilon(\lambda)\equiv\varepsilon'(\lambda)$), and all higher-order statistics are recovered exactly.}
    \label{fig:3state}
\end{figure}

Let us now focus on another paradigmatic scenario representing a building block for several out-of-equilibrium systems, ranging from molecular motors \cite{mallick2007nonequilibrium,teza2020exact} to the activity of neuronal cells \cite{grandpre2024direct}: a three state system coarse-grained into an effective two state system through a lumping of two states (Fig. \ref{fig:3state}a).
With respect to the two previous examples, this one gives the possibility to follow the lumping procedure with more analytical insight. It also shows that in general collapsing loops does not guarantee full preservation of the EPR spectrum.

In the underlying Markovian system, every site is characterized by a Markovian waiting time distribution $\psi_i(t)=A^{-1}_i e^{-t/A_i}$ with average waiting time $A_i=(\sum_{j} W_{ji})^{-1}$.
This allows for a direct calculation of the SCGF of the EPR $\varepsilon(\lambda)$ as the dominant eigenvalue of the tilted evolution operator for the EPR (see the SM \cite{SM}).
The average EPR for this system amounts to:
\begin{equation}\label{eq:avg_EPR_3state}
    \left< \sigma \right> = \frac{W_{12}W_{23}W_{31}-W_{13}W_{32}W_{21}}{\mathcal{Z}}\log \frac{W_{12}W_{23}W_{31}}{W_{13}W_{32}W_{21}}
\end{equation}
where $\mathcal{Z}=\sum_{i>j}(A^{-1}_i A^{-1}_j-W_{ij}W_{ji})$ is a normalizing factor.
From this expression we can appreciate how $\left<\sigma\right>$ is determined by the product of clockwise and counterclockwise transitions and that equilibrium is restored if and only if $W_{12}W_{23}W_{31}=W_{13}W_{32}W_{21}$.

We now coarse-grain the system by lumping together states $2$ and $3$ into a state $4$.
In this case it is possible to directly rewrite the simple diffusion equations of the system in terms of $P_1(t)$ and $P_4(t)=P_2(t)+P_3(t)$, which obviously satisfies $P_4(t)=1-P_1(t)$.
The resulting system of equations is reported in \cite{SM} and show a (Laplace transformed) waiting time distribution
\begin{equation} \textstyle \label{eq:psi4_laplace}
    \tilde{\psi}_4(u)=\frac{W_{21}[W_{13}W_{32}+W_{12}(A_3^{-1}+u)]+W_{31}[W_{12}W_{23}+W_{13}(A_2^{-1}+u)]}{A_1^{-1}[(A_2^{-1}+u)(A_3^{-1}+u)-W_{23}W_{32}]}
\end{equation}
in the lumped state, while the distribution for state $1$ remains Markovian.
Based on this construction of $\tilde{\psi}_4(u)$, 
it becomes straightforward to characterize all transitions involving state $1$ and the lumped state $4$ and the entropic contributions associated to them (see \cite{SM} for details).
So, as in the previous example, we can write a system of equations for the simultaneous evolution of entropy and displacement in the lumped state:
\begin{widetext}
\begin{equation}\label{eq:3state_lumped_GME}
{\small
    \begin{cases}
    \begin{split} \textstyle
        (u+A^{-1}_1)\tilde{P}_1(u,S) -\delta_{S,0}= \frac{u\tilde{B}(u)\tilde{\psi}_2(u)\tilde{\psi}_3(u)}{A_1^{-1}A_2^{-1}A_3^{-1}(1-\tilde{\psi}_4(u))} \Big\{ W_{21}\big[ W_{13}W_{32}\tilde{P}_4(u,S-\log \frac{W_{13}W_{32}}{W_{23}W_{31}})+W_{12}(A_3^{-1}+u)\tilde{P}_4(u,S-\log \frac{W_{12}}{W_{21}})\big] \\ \textstyle
        + W_{31}\big[W_{12}W_{23}\tilde{P}_4(u,S-\log \frac{W_{12}W_{23}}{W_{32}W_{21}})+W_{13}(A_2^{-1}+u)\tilde{P}_4(u,S-\log \frac{W_{13}}{W_{31}})\big] \Big\}
        \end{split}  \\
        \frac{u}{1-\tilde{\psi}_4(u)} \tilde{P}_4(u,S) = W_{21} \tilde{P}_1(u,S-\log \frac{W_{21}}{W_{12}})+W_{31} \tilde{P}_1(u,S-\log \frac{W_{31}}{W_{13}})
    \end{cases}
}
\end{equation}
\end{widetext}
where $\tilde{B}(u)=\frac{A_2^{-1}A_3^{-1}}{A_2^{-1}A_3^{-1}-W_{23}W_{32}\tilde{\psi}_2(u)\tilde{\psi}_3(u)}$ and we assumed the system was initialized in state $1$ with total entropy $S=0$ at $t=0$.
Summing over all entropy values $S$ one is able to restore the equations regulating the sole dynamics on the lumped network.

Reducing the system by substituting $\tilde{P}_4$ into the first equation yields a closed expression for $\tilde{P}_1$, from which one can derive the SCGF $\varepsilon'(\lambda)$ for the EPR in the lumped system. The analytic form of $\varepsilon'(\lambda)$ is lengthy and reported in the Supplemental Material~\cite{SM}. Unlike the uniform one-dimensional ring, for arbitrary transition rates $W_{ij}$ the lumped SCGF from Eqs.~\ref{eq:3state_lumped_GME} differs from that of the original process. Nevertheless, a direct calculation of the mean EPR, $\langle \sigma' \rangle = \partial_\lambda \varepsilon'(\lambda)\vert_{\lambda=0}$, agrees perfectly with Eq.~\ref{eq:avg_EPR_3state}, showing that lumping preserves the mean EPR and correctly captures the distance from equilibrium. Such fidelity is particularly relevant for molecular motors, enabling reliable estimates of ATP-consumption efficiency~\cite{mallick2007nonequilibrium, teza2020exact}.

Figure~\ref{fig:3state} illustrates a concrete example with rates $W_{12}=3$, $W_{23}=2$, $W_{31}=1$, $W_{13}=2$, $W_{32}=1$, and $W_{21}=2$. As anticipated, $\varepsilon$ and $\varepsilon'$ differ, but share the same derivative at $\lambda=0$ (Fig.~\ref{fig:3state}b). The original SCGF is symmetric about $\lambda=-1/2$, satisfying the Gallavotti--Cohen (GC) symmetry $\varepsilon(\lambda-1)=\varepsilon(-\lambda)$~\cite{lebowitz1999gallavotti}, which implies the fluctuation theorem:
\begin{equation}\label{eq:fluctuation_theorem}
   \frac{p(S/t=\sigma)}{p(S/t=-\sigma)} = e^{t\sigma}.
\end{equation}
In the $t\to\infty$ limit, $P(S,t)\sim t^{-1}p(S/t=\sigma)\sim e^{-t I(\sigma)}$, where the rate function $I(\sigma)$ follows from the G\"artner--Ellis theorem~\cite{gartner1977on,ellis1984large}, $I(\sigma)=\sup_{\lambda\in\mathbb{R}}\{\lambda\sigma-\varepsilon(\lambda)\}$, which directly yields $I(\sigma)-I(-\sigma)=-\sigma$, confirming the fluctuation theorem.

The lumped SCGF $\varepsilon'(\lambda)$ does not preserve the Gallavotti--Cohen (GC) symmetry (Fig.~\ref{fig:3state}c). Although it may appear to obey a shifted relation $\varepsilon'(\lambda-\lambda_0)=\varepsilon(-\lambda)$ for some $\lambda_0\neq1$, this is not generally true. The loss of GC symmetry signals a breakdown of the fluctuation theorem, producing a nonlinear dependence on $\sigma$ in Eq.~\ref{eq:fluctuation_theorem}. This deviation, evident in Fig.~\ref{fig:3state}d, mirrors experimental observations that have motivated studies of GC-symmetry breaking~\cite{mehl2012role, rahav2007fluctuation, puglisi2010entropy, esposito2012stochastic, bo2014entropy}. Unlike previous perturbative approaches requiring strong timescale separation, our lumping framework reproduces these deviations without such assumptions while preserving the correct mean EPR. This clarifies the mechanism behind experimental violations and suggests how accounting for these effects may improve inference of entropy production and information loss in coarse-grained models.

The lumping scheme can nonetheless preserve the full EPR statistics when certain symmetries are present in the microscopic dynamics, such as in a one-dimensional ring driven by a uniform force $f$. For the general three-state model, $\varepsilon'(\lambda)=\varepsilon(\lambda)$ whenever $W_{21}W_{13}=W_{31}W_{12}$, meaning the combined rate $3\!\to\!1\!\to\!2$ equals its reverse $2\!\to\!1\!\to\!3$ (see SM~\cite{SM}). In this symmetric case the GC symmetry and fluctuation theorem are fully recovered. Varying $W_{21}=\alpha W_{31}W_{12}/W_{13}$ with $\alpha>0$ systematically breaks this symmetry: it holds at $\alpha=1$, while deviations alter the fluctuation statistics, yielding $\mathrm{Var}'[\sigma]\neq\mathrm{Var}[\sigma]$ (Fig.~\ref{fig:3state}e).
For $\alpha \neq 1$, Fig.~\ref{fig:3state}e shows instances where $\mathrm{Var}'[\sigma] < \mathrm{Var}[\sigma]$, which can violate the thermodynamic uncertainty relation (TUR) $\mathrm{Var}[\sigma]\geq 2k_b\left<\sigma\right>$ setting a bound to the EPR fluctuation's amplitude~\cite{gingrich2016dissipation}.
The TUR is generally valid for overdamped systems; violations or extensions have been reported in underdamped dynamics~\cite{van2019uncertainty,fischer2018large,pietzonka2022classical}, time-dependent or transient driving~\cite{koyuk2020thermodynamic,macieszczak2018unified,dieball2023direct,liu2020thermodynamic,dechant2018current,koyuk2019operationally}, discrete-time dynamics~\cite{proesmans2017discrete}, and strongly correlated systems~\cite{grandpre2021entropy}.
This example provides a clear understanding of the effects of coarse-graining on TUR, underlining how the memory effects introduced by the impossibility of resolving between two states can be directly responsible for a violation on these bounds.

In this work, we showed that lumping in special cases, such as the periodic ring, or the periodic ring with secondary loops (see Appendix) act an exact coarse-graining that reproduces the microscopic entropy-production statistics~\cite{teza2020exact}. More generally, it fails to do so, breaking the Gallavotti--Cohen symmetry and other fluctuation relations~\cite{lebowitz1999gallavotti} even when the underlying dynamics obeys them.
The full success of lumping may be conditioned by a variety of algebraic or topological features of the microscopic dynamics. These results clarify how experimentally observed violations of fluctuation relations—such as Gallavotti–Cohen symmetry and thermodynamic uncertainty bounds—can emerge purely from coarse-graining, providing a principled interpretation of irreversibility
measurements in micro- and nanoscale experiments~\cite{mehl2012role,battle2016broken,diterlizzi2024variance}. Future work will explore how these limits of exact coarse-graining constrain thermodynamic inference in reduced descriptions of nonequilibrium systems~\cite{amann2010communications,wu2025parameter,martinez2019inferring, van2023time, pietzonka2024thermodynamic, grandpre2024direct}.

\begin{acknowledgments}
We thank William Bialek, Matteo Ciarchi, Ivan Di Terlizzi, and Qiwei Yu for useful discussions and remarks. G. T. acknowledges support by the Max Planck Society. T.G. acknowledges support and stimulating discussions from the National Institute for Theory and Mathematics in Biology (NITMB).
\end{acknowledgments}









\bigskip
\begin{appendix}

\begin{figure}
    \centering
    \includegraphics[width=\linewidth]{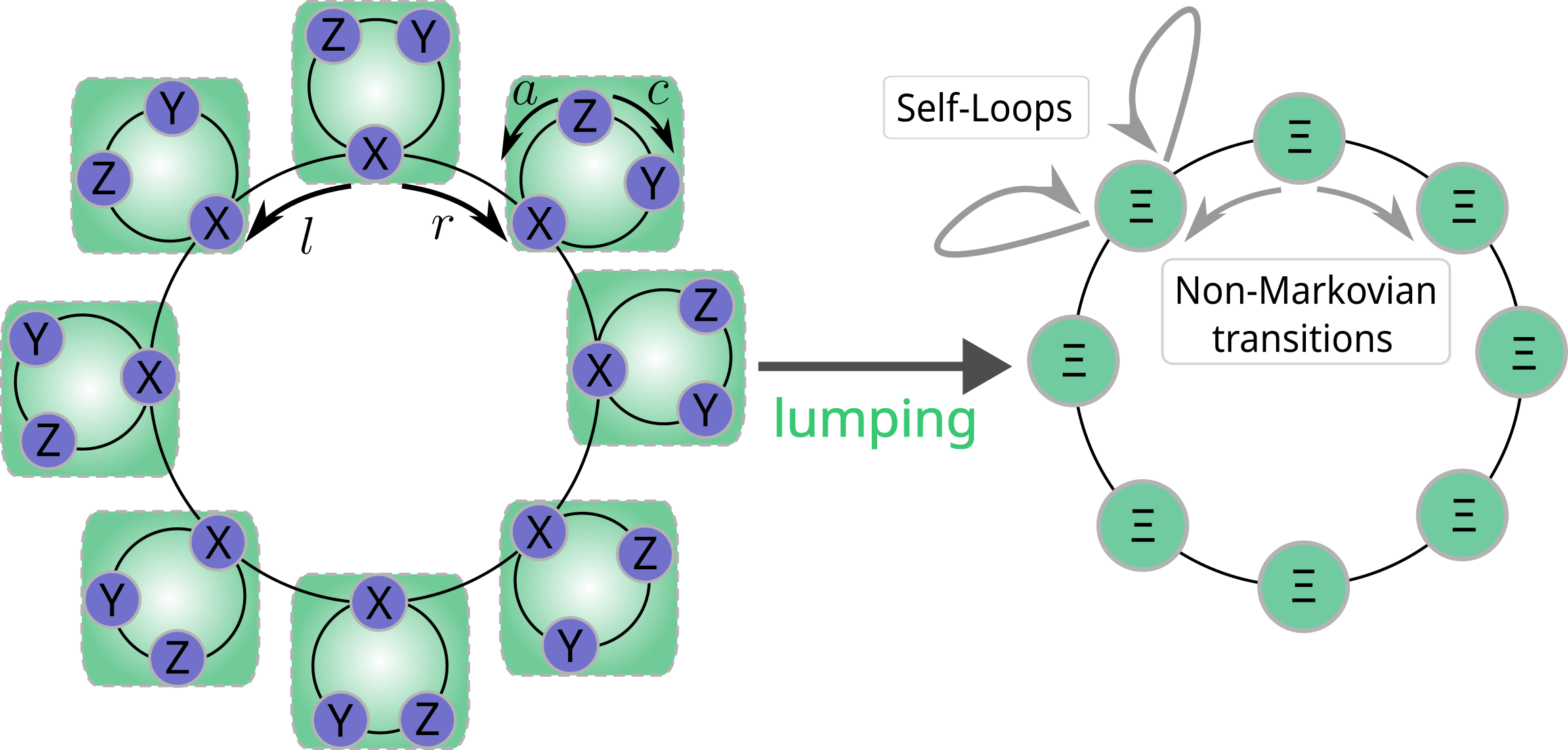}
    \caption{Sketch illustrating the lumping procedure of a secondary loops in a linear network.
    Every secondary loop is composed by a triplet of $(X,Y,Z)$ states, which are coarse-grained together in a single lumped state $\Xi$.
    The lumping results in a non-Markovian dynamics in which we are able to identify "self-loop" transitions of the lumped state to itself.
    A proper accounting of these transitions is what enables us to retain a completely exact description of the EPR in the lumped dynamics.}
    \label{figAP:secondary_loops_lumping}
\end{figure}

\renewcommand{\theequation}{A\arabic{equation}}
\setcounter{equation}{0}
\noindent \textit{Appendix A: Lumping of secondary loops}.
We consider a system composed of a main chain of $N$ sites (which we identify with the letter $X$) with hopping rates $r$ ($l$) to the right (left).
Additional secondary loops are attached to each $X$ site with two additional $Y$ and $Z$ states.
Within these secondary loops, rates $c$ and $a$ regulate clockwise and counterclockwise transitions, respectively (see Fig. \ref{figAP:secondary_loops_lumping}).
Identifying with $P_i^{\{X,Y,Z\}}(S,t)$ the probability of being in each of the three kinds of states at the $i-th$ position on the main ring, the evolution is regulated by the following master equation:
\begin{equation}\label{eqAP:second_loops_laplace_Markov}
{\small
\left\{
\begin{aligned}
u\tilde{P}_i^X(u,S) &= r\,\tilde{P}_{i-1}^X\!\left(u,S-\log\frac{r}{l}\right)
   + l\,\tilde{P}_{i+1}^X\!\left(u,S-\log\frac{l}{r}\right) \\
&\quad + a\,\tilde{P}_{i}^Z\!\left(u,S-\log\frac{a}{c}\right)
   + c\,\tilde{P}_{i}^Z\!\left(u,S-\log\frac{c}{a}\right) \\
&\quad + (r+l+a+c)\tilde{P}_i^X(u,S) + \delta_{i,0} \\[1em]
u\tilde{P}_i^Y(u,S) &= a\,\tilde{P}_{i}^X\!\left(u,S-\log\frac{a}{c}\right)
   + c\,\tilde{P}_{i}^Z\!\left(u,S-\log\frac{c}{a}\right) \\
&\quad + (a+c)\tilde{P}_i^Y(u,S) \\[1em]
u\tilde{P}_i^Z(u,S) &= c\,\tilde{P}_{i}^X\!\left(u,S-\log\frac{c}{a}\right)
   + a\,\tilde{P}_{i}^Y\!\left(u,S-\log\frac{a}{c}\right) \\
&\quad + (a+c)\tilde{P}_i^Y(u,S)
\end{aligned}
\right.
}
\end{equation}
where we introduced the Laplace transforms $\tilde{P}_{i}^{\{X,Y,Z\}}(S,u)=\int_0^{\infty}dt e^{tu} P_{i}^{\{X,Y,Z\}}(S,t)$ and assumed initial condition on the $X$ state at $i=0$.
Upon exact decimation \cite{teza2020exact,teza2020thesis}, one can find the SCGF function of the original process to be the solution of the third-order polynomial equation
\begin{equation}\label{eqAP:second_loops_SCGF_3rd_order_poly}
    \varepsilon(\lambda)^3 + \alpha(\lambda) \varepsilon(\lambda)^2 +  \beta(\lambda) \varepsilon(\lambda) +  \gamma(\lambda) = 0 .
\end{equation}
The explicit expressions of the coefficients are reported in the Supplemental Material \cite{SM}.
The SCGF provides us with direct access to all the cumulants of the EPR upon differentiation with respect to $\lambda$ in 0, yielding the following average EPR:
\begin{equation}
    \left< S/t \right>=(a-c) \log \frac{a}{c} + 3(r-l) \log \frac{r}{l} .
\end{equation}

Lumping together the secondary loops consisting into a triplet of $(X,Y,Z)$ states into a single effective state $\Xi$ provides us with a coarse-grained non-Markovian system evolving on the main loop (see sketch of Fig. \ref{figAP:secondary_loops_lumping}).
When the particle is hopping inside the secondary loop, it can produce entropy (in the case $a\neq c$) by performing a full revolution in either direction.
In the evaluation of the waiting time distribution $\tilde{\Psi}(u)$ of the lumped state $\Xi$, we will therefore account a full revolution  in the secondary loop (in either direction) as a "self-transition" to not lose the contribution to the total EPR.

The full derivation end expression of the waiting time distribution $\tilde{\Psi}(u)$ is reported in the Supplemental Material~\cite{SM}.
Here we report the full form of the associated memory kernel $\tilde{W}(u)=s\tilde{\Psi}(u)/(1-\tilde{\Psi}(u))$:
\begin{equation}
    \tilde{W}(u)=
    \frac{a^3+c^3+(r+l)\left(a^2+ac+c^2+2(a+c)u+u^2\right)}{3(a^2+ac+c^2)+3(a+c)s+s^2}
\end{equation}
Here, we can immediately distinguish how each term in the memory kernel is associated with different trajectories, and hence, each differentiate the contribution to the total EPR.
The terms $\propto a^3$ and $\propto c^3$ account for a full revolution performed in either direction, while the terms $\propto r$ and $\propto l$ account for the non-Markovian transitions resulting in the main loop (see right sketch in Fig.~\ref{figAP:secondary_loops_lumping}).

Following this reasoning we introduce the kernels $\tilde{W}_{\{r,l,a,c\}}(u)$
such that $\tilde{W}(u)\equiv \tilde{W}_r(u)+ \tilde{W}_l(u)+ \tilde{W}_a(u)+ \tilde{W}_c(u)$.
This way, we can now write a generalized master equation for the probability $P^{\Xi}_i(t,S)$ of being in the lumped state$ \Xi$ at position $i$ at a given time $t$ system that properly accounts for the total produced entropy $S$:
\begin{align}\label{eqAP:second_loops_GME_laplace}\nonumber
    (s+\tilde{W}(u))\tilde{P}_i^\Xi(u,S) -\delta_{i,0} &=
    \tilde{W}_r(u)\tilde{P}_{i-1}^\Xi(u,S-\log \frac{r}{l})+ \\ \nonumber
    &\tilde{W}_l(u)\tilde{P}_{i+1}^\Xi(u,S-\log \frac{l}{r})+ \\ \nonumber
    &\tilde{W}_a(u)\tilde{P}_{i+1}^\Xi(u,S-3\log \frac{a}{c})+ \\
    &\tilde{W}_c(u)\tilde{P}_{i+1}^\Xi(u,S-3\log \frac{c}{a})
\end{align}
We underline how summing over the entropy would provide us with the expected equation for $\tilde{P}^{\Xi}_i(u)=\sum_S \tilde{P}^{\Xi}_i(u,S)$ describing the spatial evolution on the main loop of the lumped system.

Inverse Laplace transforming Eq.~\ref{eqAP:second_loops_GME_laplace} allows us to obtain a third order differential equation for the cumulant generating function of the EPR of the lumped state $G'(\lambda,t)=\sum_{i,S}P^{\Xi}_ie^{\lambda S} (t,S)$.
In the approaching to the stationary state, the large deviation principle ensures that $G'(\lambda,t)\to e^{\varepsilon'(\lambda)t}$ with $\varepsilon'(\lambda)$ SCGF of the EPR of the lumped system.
Doing so, one recovers precisely the polynomial equation for the original SCGF (Eq.~\ref{eqAP:second_loops_SCGF_3rd_order_poly}), implying that the lumped description captures \textit{exactly} the entire EPR distribution of the original process.

This example shows on one hand how our lumping procedure allows to preserve all the entropy production also in an explicit case in which entire loops producing entropy are erased, while also showacasing a scenario in which the lumping and decimation procedure yield the same kind of coarse-grained non-Markovian dynamics.
This coincidence is due to the circumstance that in a decimated situation the arrival to/departure from an $X$ state precisely monitors also the arrival to/departure from a lumped state.

\end{appendix}

\onecolumngrid

\bigskip
\hrule
\bigskip

\setcounter{equation}{0}
\renewcommand{\theequation}{S\arabic{equation}}

\begin{center}
    {\LARGE \textbf{Supplemental Material (SM)}}
\end{center}

In this supplemental material (SM) we discuss the details of the calculations and results presented in the main text.

\section{Lumping of a 1D ring}

Let us consider the paradigmatic example of a 1-dimensional periodic discrete lattice in which a particle hops with some rate $r$ to the right and $l$ to the left with Markovian dynamics.
The probability of being on a site $i$ at a generic time $t$ with an overall accumulated entropy production $S$ evolves according to the master equation \cite{teza2020exact}
\begin{equation} \label{eqSM:microscopic_ME}
    [r+l+\partial_t]P_i(t,S)=r P_{i-1}(t,S-\log r/l)+lP_{i+1}(t,S-\log l/r) .
\end{equation}
In such a model, the waiting time distribution (i.e. the distribution of the time spent by a walker on each site of the lattice) is an exponential distribution $\psi(t)=(r+l)e^{-(r+l)t}$, where the characteristic time is set by the jump rates.

Our goal is to describe the evolution of the lumped system (as depicted in the sketch of Fig. \ref{fig:lumping_sketch}a of the main text), therefore it is convenient to switch to Laplace transforms in the time domain, where we define $\tilde{f}(u)=\int_0^{+\infty} dt f(t) e^{-ut}$ for some generic function $f(t)$ defined for $t\in[0;+\infty)$.
The Laplace transform of the exponential time distribution can be easily verified to read
\begin{equation}
    \psi(u)=\frac{r+l}{u+r+l} .
\end{equation}
while the waiting time distribution $\Psi$ of the lumped state can be expressed as a convolution of the microscopic $\psi$ which, in Laplace domain reads:
\begin{eqnarray}\label{eqSM:Psi_1D_lump}
    \tilde{\Psi}(u) &&=\frac{\tilde{B}(s)\tilde{\psi}^2(u)}{(r+l)^3} \left[ \frac{2 rl (r+l)}{\tilde{\psi}(u)}+r^3+l^3 \right] = \\ \nonumber
    &&= \frac{1}{(u+r+l)^2-rl} \frac{2 rl(r+l+u)+r^3+l^3}{r+l} 
\end{eqnarray}
where
\begin{equation}
    \tilde{B}(u)=\frac{1}{1-\frac{rl}{(r+l)^2}\tilde{\psi}(u)^2}=\frac{(u+r+l)^2}{(u+r+l)^2-rl}
\end{equation}
is the 'bouncing' waiting time distribution accounting for an indefinite number of jumps withing the two microstates composing the lumped state.
The other terms in Eq. \ref{eqSM:Psi_1D_lump} are directly related to a specific way of entering and exiting the lumped state: the $r^3$ ($l^3$) contribution is associated with the probability of entering the state from the left (right) and exiting to the right (left), while the term $\propto 2rl$ accounts for trajectories entering and exiting the lumped state from the same side. 
An explicit expression in time coordinates can be evaluated analytically by inverse-Laplace transforming $\tilde{\Psi}(u)$, resulting into:
\begin{equation}
    \Psi(t)=\frac{\left(l^3+r^3\right) \sinh \left(t \sqrt{l r}\right)+2 (l r)^{\frac{3}{2}} \cosh \left(t \sqrt{l r}\right)}{\sqrt{l r} (l+r)} e^{-t (l+r)} 
\end{equation}

The memory kernel associated with this waiting time distribution $\Psi$ can be expressed through a straightforward relation in Laplace coordinates \cite{klafter1980derivation}:
\begin{equation}
    \tilde{W}(u)=\frac{u\tilde{\Psi}(u)}{1-\tilde{\Psi}(u)}=\frac{2rl(r+l+u)+r^3+l^3}{l^2+r^2+(l+r)(l+r+u)} .
 \end{equation}
This enables us to write a generalized master equation to describe the dynamics (not accounting for the entropy production) of the lumped system as:
\begin{equation}\label{eqSM:GME_1D_ring_laplace_no_ent}
    [u+\tilde{W}(u)]\tilde{P}_I(u)-\tilde{P}_I(t=0)=\tilde{W}_r(u)\tilde{P}_{I-1}(u)+\tilde{W}_l(u)\tilde{P}_{I+1}(u)
\end{equation}
where used capital $I$ to refer to the position index of lumped states and we introduced the right and left transition kernels:
\begin{eqnarray}
    \tilde{W}_{r}(u)&= \frac{u}{1-\tilde{\Psi}(u)} \frac{\tilde{B}(u)\tilde{\psi}^2(u)}{(r+l)^3}  \left[ \frac{ rl (r+l)}{\tilde{\psi}(u)}+r^3 \right] \\ \nonumber
    \tilde{W}_{l}(u)&= \frac{u}{1-\tilde{\Psi}(u)} \frac{\tilde{B}(u)\tilde{\psi}^2(u)}{(r+l)^3}  \left[ \frac{ rl (r+l)}{\tilde{\psi}(u)}+l^3 \right]
\end{eqnarray}
such that $\tilde{W}_r(u)+\tilde{W}_l(u)\equiv \tilde{W}(u)$.

Substituting the kernel expressions in equation, one obtains the following generalized master equation:
\begin{equation}
    u \left(1+\frac{\tilde{\Psi}(u)}{1-\tilde{\Psi}(u)} \right) (\tilde{P}_I(u) - P_I(t=0))= \frac{u}{1-\tilde{\Psi}(u)} \frac{\tilde{B}(u)\psi^2(u)}{(r+l)^3}
    \left[ \left( \frac{ rl (r+l)}{\tilde{\psi}(u)}+r^3 \right) \tilde{P}_{I-1}(u)+
    \left( \frac{ rl (r+l)}{\tilde{\psi}(u)}+l^3 \right) \tilde{P}_{I+1}(u)
    \right]
\end{equation}
where $P_I(t=0)$ is the initial condition of the problem.
After some manipulation, the above equation can be simplified to
\begin{equation}
    \tilde{P}_I(u) - \tilde{P}_I(t=0) = \frac{\tilde{B}(u)\tilde{\psi}^2(u)}{(r+l)^3}
    \left[ \left( \frac{ rl (r+l)}{\tilde{\psi}(u)}+r^3 \right) \tilde{P}_{I-1}(u)+
    \left( \frac{ rl (r+l)}{\tilde{\psi}(u)}+l^3 \right) \tilde{P}_{I+1}(u) \right]
\end{equation}

Let us now keep into account the entropy contribution coming for each transition, which will depend on the entrance point (from the left or the right) in the lumped state.
Doing so one gets:
\begin{eqnarray}
    \tilde{P}_I(u,S) - P_I(t=0) &= \frac{\tilde{B}(u)\tilde{\psi}^2(u)}{(r+l)^3}
    \Big[  \frac{ rl (r+l)}{\tilde{\psi}(u)}\tilde{P}_{I-1}(u,S-\log r/l)+r^3 \tilde{P}_{I-1}(u,S-2\log r/l)+ \\ \nonumber
    &+ \frac{ rl (r+l)}{\tilde{\psi}(u)}\tilde{P}_{I+1}(u,S-\log l/r)+l^3  \tilde{P}_{I+1}(u,S-2\log l/r) \Big]
\end{eqnarray}
Now, we substitute the explicit expressions of $\tilde{B}(u)$ and $\tilde{\psi}(u)$ and bring all the terms containing $u$ to the numerator to highlight time deravites in Laplace coordinates.
Doing so we get:
\begin{eqnarray}
    [(u+r+l)^2-rl](\tilde{P}_I(u,S)-P_I(t=0)) &=\frac{1}{r+l} \big[ rl(u+r+l) \tilde{P}_{I-1}(u,S-\log r/l)+r^3\tilde{P}_{I-1}(u,S-2\log r/l)  \\ \nonumber
     &+rl(u+r+l) \tilde{P}_{I+1}(u,S-\log l/r) + l^3 \tilde{P}_{I+1}(u,S-2\log l/r) \big]
\end{eqnarray}
which can be seen to provide the dynamics of Eq. \ref{eqSM:GME_1D_ring_laplace_no_ent} upon summation on all possible values of the entropy production $S$.
Finally, one can inverse-Laplace transform the above equation to go back to the time domain obtaining:
\begin{eqnarray} \label{eqSM:1D_lumped_GME}
    [\partial_t^2 +2(r+l)\partial_t+r^2+rl+l^2]P_I(t,S) &=\frac{1}{r+l} \big[ rl(\partial_t+r+l) P_{I-1}(t,S-\log r/l)+r^3P_{I-1}(t,S-2\log r/l)  \\ \nonumber
     &+     rl(\partial_t+r+l) P_{I+1}(t,S-\log l/r) + l^3 P_{I+1}(t,S-2\log l/r) \big]
\end{eqnarray}

To obtain the overall entropy production accumulated at a given time $t$ associated with this process, one needs to sum over all positions $I$ introducing a $P(t,S):=\sum_I P_I(t,S)$.
The resulting equation reads:
\begin{eqnarray}
    [\partial_t^2 +2(r+l)\partial_t+r^2+rl+l^2]P(t,S) &=\frac{1}{r+l} \big[ rl(\partial_t+r+l) P(t,S-\log r/l)+r^3P(t,S-2\log r/l)  \\ \nonumber
     &+     rl(\partial_t+r+l) P(t,S-\log l/r) + l^3 P(t,S-2\log l/r) \big]
\end{eqnarray}
Introducing the generating function of the entropy production $G(\lambda,t)=\sum_S e^{\lambda S} P(t,S)$ one can see (multiply by $e^{\lambda S}$ and sum over all $S$) that it satisfies the following second order differential equation:
\begin{eqnarray}
    [\partial_t^2 +2(r+l)\partial_t+r^2+rl+l^2]G(\lambda,t) &=\frac{1}{r+l} \big[ rl(\partial_t+r+l) e^{\lambda\log r/l}+r^3e^{2\lambda\log r/l}  \\ \nonumber
     &+     rl(\partial_t+r+l) e^{\lambda\log l/r} + l^3 e^{2\lambda\log l/r} \big] G(\lambda,t)
\end{eqnarray}
which simplifies to
\begin{eqnarray} \label{eqSM:G_entropy_diff_eq}
    [\partial_t^2 +\frac{2(r+l)^2-rl(e^{\lambda\log r/l}+e^{\lambda\log l/r})}{r+l}\partial_t ]G(\lambda,t) &=
     \big[ - (r^2+rl+l^2) +\frac{r^3}{r+l}e^{2\lambda\log r/l}  \\ \nonumber
     &+    rl(e^{\lambda\log r/l}+e^{\lambda\log l/r}) + \frac{l^3}{r+l} e^{2\lambda\log l/r} \big] G(\lambda,t)
\end{eqnarray}
The scaled cumulant generating function (SCGF) is related with the time-asymptotic form of the generating function $G(\lambda,t)\sim e^{\varepsilon(\lambda)t}$, or, more formally by the following limit \cite{touchette2009large}:
\begin{equation}
\varepsilon(\lambda)=\lim_{t\to\infty} \frac{\log G(\lambda,t)}{t}     .
\end{equation}
Substituting this ansatz in Eq. \ref{eqSM:G_entropy_diff_eq} provides us with the following second order polynomial equation for $\varepsilon(\lambda)$:
\begin{eqnarray}
    &\varepsilon^2(\lambda) +\frac{2(r+l)^2-rl(e^{\lambda\log r/l}+e^{\lambda\log l/r})}{r+l}\varepsilon(\lambda) + \\ \nonumber
    &+ r^2+rl+l^2 - rl(e^{\lambda\log r/l}+e^{\lambda\log l/r})-\frac{r^3}{r+l}e^{2\lambda\log r/l} - \frac{l^3}{r+l} e^{2\lambda\log l/r} =0
\end{eqnarray}
Which can be straightforwardly solved to obtain the SCGF from its dominant root:
\begin{equation}
    \varepsilon(\lambda)=-(r+l)+r e^{\lambda \log r/l}+l e^{\lambda \log l/r} .
\end{equation}
This expression coincides with the SCGF of the original (non-lumped) system, implying that the equation we wrote for the lumped system captures the whole distribution of entropy production of the original microscopic system.

\subsection{Markovianization of the 1D ring}
We can ask ourselves if one can restore a Markovian description of the lumped system while being faithful to the original entropy production.
This procedure is performed in the same spirit of the "markovianization" procedure presented in \cite{teza2020thesis}.
The idea is to simply drop higher order derivatives in the generalized master equation for the lumped system and properly expand the gradient terms corresponding to time derivatives not centered in the site of interest.
The latter terms can be rewritten as
\begin{equation}
    P_{I \pm 1}(t,S)= P_I(t,S) + L\frac{P_{I\pm1}(t,S)-P_I(t,S)}{L}  
\end{equation}
where $L$ is the lattice spacing.
The second term can be interpreted as the discrete analogue of a space derivative, so that in our approximation we can ignore it as subleading whenever it's derived in time implying
\begin{equation}
    \partial_t P_{I \pm 1}(t,S)= \partial_t P_I(t,S) + L\frac{\partial_t [P_{I\pm1}(t,S)- P_I(t,S)]}{L}\sim \partial_t P_I(t,S)
\end{equation}
Following these approximations Eq. \ref{eqSM:1D_lumped_GME} becomes:
\begin{eqnarray}
    [2 \frac{r^2+rl+l^2}{r+l}\partial_t+r^2+rl+l^2]P_I(t,S) &=\frac{1}{r+l} \big[ rl(r+l) P_{I-1}(t,S-\log r/l)+r^3P_{I-1}(t,S-2\log r/l)  \\ \nonumber
     &+     rl(r+l) P_{I+1}(t,S-\log l/r) + l^3 P_{I+1}(t,S-2\log l/r) \big]
\end{eqnarray}
Proceeding as above to evaluate the SCGF of this approximated system we get:
\begin{equation}
    \varepsilon'(\lambda)=-\frac{r+l}{2}+\frac{1}{2(r^2+rl+l^2)} \left[ r^3e^{2\lambda\log r/l} + l^3e^{2\lambda\log l/r} +rl(r+l) (e^{\lambda\log r/l}+e^{\lambda\log l/r})\right]
\end{equation}
Straightforward calculations show that
\begin{equation}
    \partial_\lambda \varepsilon(\lambda)|_{\lambda=0}=(r-l)\log\frac{r}{l}
\end{equation}
meaning that the average entropy production matches that of the original system.
This does not hold for the second (or higher) cumulants.

\section{Lumping of a 3-state system into an effective 2-state system}

The dynamics of a fully connected three-state Markovian network is described by the following system of master equations:
\begin{eqnarray}\label{eqSM:3_state_Markov}
    \partial_t P_1(t)  &= -\frac{1}{A_1}P_1(t)+W_{12}P_2(t)+W_{13}P_3(t) \\ \nonumber
    \partial_t  P_2(t) &= -\frac{1}{A_2}P_2(t)+W_{21}P_1(t)+W_{23}P_3(t) \\ \nonumber
    \partial_t  P_3(t) &= -\frac{1}{A_3}P_3(t)+W_{31}P_1(t)+W_{32}P_2(t)
\end{eqnarray}
where $W_{ij}$ are the transition rates from state $j$ to state $i$ and $A_i=(\sum_{j} W_{ji})^{-1}$ is the average waiting time for the state $i$, regulated by the exponential waiting time distribution $\psi(t)=A_i^{-1}e^{-t/A_i}$.

For convenience, let us express below here the steady state configuration (marked with an $^*$), obtained by setting to zero all time derivatives.
\begin{equation}
    \begin{cases}
        P_1^* &\propto A_2^{-1}A_3^{-1}-W_{23}W_{32} \\
        P_2^* &\propto A_1^{-1}A_3^{-1}-W_{13}W_{31} \\
        P_3^* &\propto A_1^{-1}A_2^{-1}-W_{12}W_{21} 
    \end{cases}
\end{equation}
The normalization coefficient is found imposing $\sum_i P_i^*=1$, and is found to be
\begin{equation}
    \mathcal{Z}=A_2^{-1}A_3^{-1}+A_1^{-1}A_3^{-1}+A_1^{-1}A_2^{-1}-W_{23}W_{32}-W_{13}W_{31}-W_{12}W_{21}
\end{equation}
The entropy production rate in the steady state can be expressed as $\left< S/t \right> = \sum_{ij} W_{ji} P^*_i \log W_{ji}/W_{ij}$ \cite{schnakenberg1976network}, which, for this three-state systems reads:
\begin{equation}\label{eqSM:entr_prod_rate}
    \left< S/t \right> = \frac{W_{12}W_{23}W_{31}-W_{13}W_{32}W_{21}}{\mathcal{Z}}\log \frac{W_{12}W_{23}W_{31}}{W_{13}W_{32}W_{21}} .
\end{equation}

\subsection{Lumped equations of the dynamics}
Switching to Laplace space, the system of Eq. \ref{eqSM:3_state_Markov} becomes
\begin{eqnarray}\label{eqSM:3_state_laplace_Markov}
    u \tilde{P}_1(u) - 1 &= -\frac{1}{A_1}\tilde{P}_1(u)+W_{12}\tilde{P}_2(u)+W_{13}\tilde{P}_3(u) \\ \nonumber
    u \tilde{P}_2(u) &= -\frac{1}{A_2}\tilde{P}_2(u)+W_{21}\tilde{P}_1(u)+W_{23}\tilde{P}_3(u) \\ \nonumber
    u \tilde{P}_3(u) &= -\frac{1}{A_3}\tilde{P}_3(u)+W_{31}\tilde{P}_1(u)+W_{32}\tilde{P}_2(u)
\end{eqnarray}
where we used as $P_i(t=0)=\delta_{i,1}$ as initial condition.
If we are encountering a situation in which the second and third states are lumped into a single state such that $P_4(u)\equiv P_2(u)+P_3(u)$ for every $u$, we should be able to algebraically manipulate the above system into the following non-markovian system
\begin{eqnarray}\label{eqSM:3_state_Markov_CG}
    u \tilde{P}_1(u) - 1 &= -\frac{1}{A_1}\tilde{P}_1(u)+\tilde{w}_{4}(s)\tilde{P}_4(u) \\ \nonumber
    u \tilde{P}_4(u) &= -\tilde{w}_{4}(u)\tilde{P}_4(u)+\frac{1}{A_1}\tilde{P}_1(u)
\end{eqnarray}
for some memory kernel $\tilde{w}_4(u)=\frac{s\tilde{\psi}_4(u)}{1-\tilde{\psi}_4(u)}$, with $\tilde{\psi}_4(u)$ being the waiting time distribution of the lumped state.

Let us therefore perform an algebraic reduction of the system of Eq. \ref{eqSM:3_state_laplace_Markov} by solving the second and third equations for $P_2$ and $P_3$ obtaining:
\begin{eqnarray}
    \tilde{P}_2(u) &= \frac{W_{21}(u+A_3^{-1})+W_{23}W_{31}}{(u+A_2^{-1})(u+A_3^{-1})-W_{23}W_{32}}  \tilde{P}_1(u)\\ \nonumber
    \tilde{P}_3(u) &= \frac{W_{31}(u+A_3^{-1})+W_{32}W_{21}}{(u+A_2^{-1})(u+A_3^{-1})-W_{23}W_{32}}  \tilde{P}_1(u)
\end{eqnarray}
Substituting these expression into the first equation after some manipulation we get:
\begin{equation}
    \left( u+\frac{1}{A_1} \right)\tilde{P}_1(u)-1=
     \frac{W_{21}[W_{13}W_{32}+W_{12}(A_3^{-1}+u)]+W_{31}[W_{12}W_{23}+W_{13}(A_2^{-1}+u)]}{A_1^{-1}[(A_2^{-1}+u)(A_3^{-1}+u)-W_{23}W_{32}]}\frac{P_1(s)}{A_1}
    \equiv \tilde{\psi}_4(u) \frac{\tilde{P}_1(u)}{A_1}
\end{equation}
where we identified the waiting time distribution of the lumped state introduced in Eq. \ref{eq:psi4_laplace} of the main text.

Now we can see that this result coincides with the reduced system of Eq. \ref{eqSM:3_state_Markov_CG} in which the lumped state $\tilde{P}_4(u)=\tilde{P}_2(u)+\tilde{P}_3(u)$ was introduced.
Indeed, solving the first equation for $\tilde{P}_4(u)$ provides us with:
\begin{equation}
    \tilde{P}_4(u)=\frac{1}{u+\tilde{w}_4(u)}\frac{\tilde{P}_1(u)}{A_1}
\end{equation}
which, substituted in the first equations yields:
\begin{equation}
     \left( u+\frac{1}{A_1} \right)\tilde{P}_1(u)-1=\frac{\tilde{w}_4(u)}{u+\tilde{w}_4(u)}\frac{\tilde{P}_1(u)}{A_1}\equiv \tilde{\psi}_4(u) \frac{\tilde{P}_1(u)}{A_1}
\end{equation}
where we used the relation $\tilde{\psi}(u)=\tilde{w}(u)/(u+\tilde{w}(u))$ connecting an RTD with its memory kernel, proving that the reduced system is exactly described by Eq. \ref{eqSM:3_state_Markov_CG}.

\subsection{Lumped equations for the EPR}
In order to account for the contributions to the EPR in the lumped description, we need to associate each term in the waiting time distribution $\psi_4$ of the lumped state.
Indeed, while the RTDs for state $1$ is left unaltered and will retain its original Markovian character (exponential form), the one of the new lumped state can be expressed as a convolution of exponential distributions keeping into account all the possible transitions that can occur inside the lumped state before exiting.

First, we evaluate the probability of performing an indefinite number of 'bounces' between $2$ and $3$, which can be expressed as the following geometric series:
\begin{equation}
    \tilde{B}(u)\equiv \sum_{n=0}^{\infty}[A_2 A_3 W_{23}W_{32} \tilde{\psi}_2(s) \tilde{\psi}_{3}(s)]^n=
    \frac{A_2^{-1} A_3^{-1}}{A_2^{-1} A_3^{-1} -W_{23}W_{23} \tilde{\psi}_2(u) \tilde{\psi}_{3}(u)}
\end{equation}
This contribution is always present regardless of the specific way (either through a transition $1\to2$ or $1\to3$) the system enters in the lumped state, hence we will factor it out in the following reasoning.

The specific microscopic transition by which the system enters the lumped state will determine different possible trajectories.
With probability $W_{21}A_1$ the system will enter through site $2$, after which it can exit the state either by retracing its steps with a $2\to1$ transition or by closing the loop with a double $2\to3\to1$ transition.
An analogous reasoning applies for the case in which $1\to3$ is what brings the system in the lumped state.
Putting everything together, one ultimately finds the following expression for the waiting time distribution of the lumped state:
\begin{eqnarray}\label{eqSM:psi4_laplace}
    \tilde{\psi}_4(u)= \tilde{B}(u) \Big\{
   & \left[ W_{12}  + W_{13}A_3 \tilde{\psi}_3(u) W_{32}   \right]  A_2 \tilde{\psi}_2(u) W_{21}\\ \nonumber
   & + \left[ W_{13}  + W_{12}A_2 \tilde{\psi}_2(u) W_{23}   \right]  A_3 \tilde{\psi}_3(u) W_{31}
   \Big\} A_1
\end{eqnarray}
where one can recognize in the first (second) row the contributions when entering the lumped state through $2$ ($1$).
Substituting $\tilde{\psi}_{2,3}(u)=1/(1+A_{2,3}u)$ in the above formula yields precisely Eq. \ref{eq:psi4_laplace} of the main text, matching the results obtained with the algebraic lumping of the dynamics equations.

In Eq. \ref{eqSM:psi4_laplace} there are 4 contributions, each associated with a specific trajectory.
Concerning the trajectories exiting and entering the lumped state from the same side, the term $\propto W_{12}$ will provide an EPR contribution $\log W_{12}/W_{21}$, while the term $\propto W_{13}$ will contribute with $\log W_{13}/W_{31}$.
Concerning the remaining two trajectories in which the state enters and exits the lumped state from different sites, the term $\propto W_{13}W_{32}$ will provide a contribution $\log W_{13}W_{32}/W_{23}W_{31}$, while the term $\propto W_{12}W_{23}$ will contribute with $\log W_{12}W_{23}/W_{32}W_{21}$.
Putting everything together, and reminding that the transitions from state $1$ to state $4$ were left unaltered and stayed Markovian, one ultimately obtains the lumped description of Eq. \ref{eq:3state_lumped_GME} of the main text.
Reducing the system by substituting the expression of $\tilde{P}_4$ in the first equation, one obtains the following equation for the dynamics: 
\begin{equation}\label{eqSM:3state_lumped_GME}
    \begin{split} \textstyle
        (u+A^{-1}_1)\tilde{P}_1(u,S) -\delta_{S,0}= \frac{\tilde{B}(u)\tilde{\psi}_2(u)\tilde{\psi}_3(u)}{A_1^{-1}A_2^{-1}A_3^{-1}} \Big\{ W_{21}\big[ W_{13}W_{32}W_{21}\tilde{P}_1(u,S-\log \frac{W_{13}W_{32}W_{21}}{W_{12}W_{23}W_{31}})+W_{12}W_{21}(A_3^{-1}+u)\tilde{P}_1(u,S) \\ \textstyle
        + W_{13}W_{32}W_{31}\tilde{P}_1(u,S-\log \frac{W_{32}}{W_{23}})+W_{12}W_{31}(A_3^{-1}+u)\tilde{P}_1(u,S-\log \frac{W_{12}W_{31}}{W_{13}W_{21}})\big] \\ \textstyle
        + W_{31}\big[W_{12}W_{23}W_{21}\tilde{P}_1(u,S-\log \frac{W_{23}}{W_{32}})+W_{13}W_{21}(A_2^{-1}+u)\tilde{P}_1(u,S-\log \frac{W_{13}W_{21}}{W_{12}W_{31}}) \\ \textstyle
        + W_{12}W_{23}W_{31}\tilde{P}_1(u,S-\log \frac{W_{12}W_{23}W_{31}}{W_{13}W_{32}W_{21}})+W_{13}W_{31}(A_2^{-1}+u)\tilde{P}_1(u,S)\big] \Big\}
    \end{split}
\end{equation} 
which, after some manipulation, can be reduced to a third order polynomial equation in the Laplace variable $u$ corresponding to a third-order differential equation in time domain.
Introducing the EPR generator $G'(\lambda,t)=\sum_{S,I} e^{\lambda S} P_I(t,S)\sim_{t\to\infty} e^{t \varepsilon'(\lambda)}$, one ultimately obtains the third order polynomial equation
\begin{equation}\label{eqSM:SCGF_3rd_order_poly}
    \varepsilon'(\lambda)^3 + \alpha(\lambda) \varepsilon'(\lambda)^2 +  \beta(\lambda) \varepsilon'(\lambda) +  \gamma(\lambda) = 0  
\end{equation}
with coefficients
\begin{eqnarray}
    \alpha(\lambda) &&= A_1^{-1}+A_2^{-1}+A_3^{-1}\\ \nonumber
    \beta(\lambda) &&= \frac{A_1+A_2+A_3}{A_1 A_2 A_3}
    - A_1 W_{13}W_{31}\Big[W_{21}e^{\lambda \log \frac{W_{13}W_{21}}{W_{12}W_{31}}}+W_{31}\Big]
    - A_1 W_{12}W_{21}\Big[W_{31}e^{\lambda \log \frac{W_{12}W_{31}}{W_{13}W_{21}}}+W_{21}\Big]
    - W_{23}W_{32}\\ \nonumber
    \gamma(\lambda) &&= \frac{1}{A_1 A_2 A_3} - \frac{W_{23}W_{32}}{A_1}
    -  \frac{A_1 W_{12}W_{21}}{A_3} \Big[W_{21}+W_{31}e^{\lambda \log \frac{W_{12}W_{31}}{W_{13}W_{21}}}\Big]
    -  \frac{A_1 W_{13}W_{31}}{A_2} \Big[W_{31}+W_{21}e^{\lambda \log \frac{W_{13}W_{21}}{W_{12}W_{31}}}\Big] \\ \nonumber
    &&-  A_1 W_{12}W_{23}W_{31} \Big[W_{21}e^{\lambda \log \frac{W_{23}}{W_{32}}}+W_{31}e^{\lambda \log \frac{W_{12}W_{23}W_{31}}{W_{13}W_{32}W_{21}}}\Big]
    -  A_1 W_{13}W_{32}W_{21} \Big[W_{31}e^{\lambda \log \frac{W_{32}}{W_{23}}}+W_{21}e^{\lambda \log \frac{W_{13}W_{32}W_{21}}{W_{12}W_{23}W_{31}}}\Big]
\end{eqnarray}
The SCGF of the lumped system $\varepsilon'(\lambda)$ can be found as the dominant root of the cubic Eq. \ref{eqSM:SCGF_3rd_order_poly}.
This provides us with access to all the scaled cumulants of the EPR, which allows one to show that
\begin{equation}
    \partial_\lambda\varepsilon'(\lambda)|_{\lambda=0}\equiv\partial_\lambda\varepsilon(\lambda)|_{\lambda=0}
\end{equation}
implying that the average value of the EPR in the lumped description matches exactly that of the original system (reported explicitly in Eq. \ref{eqSM:entr_prod_rate}).
However, higher cumulants do not match, in agreement with the example provided in Fig. \ref{fig:3state}b of the main text (for a specific set of rates).

\section{Lumped secondary loops}
Even though the three-state system presented in the main text already serves as an example for a lumping strategy that erases a loop, we present here a more explicit case in this regard.
We consider a system composed of a main chain of $N$ sites (which we identify with the letter $X$) with hopping rates $r$ ($l$) to the right (left).
Additional secondary loops are attached to each $X$ site with two additional $Y$ and $Z$ states.
Within these secondary loops, rates $c$ and $a$ regulate clockwise and counterclockwise transitions, respectively.
Identifying with $P_i^{\{X,Y,Z\}}(S,t)$ the probability of being in each of the three kinds of states at the $i-th$ position on the main ring, the evolution is regulated by the following master equation
\begin{eqnarray}\label{eqSM:second_loops_laplace_Markov}
    [r+l+a+c+u] \tilde{P}_i^X(u,S) - \delta_{i,0} &= r \tilde{P}_{i-1}^X(u,S-\log \frac{r}{l})+l\tilde{P}_{i+1}^X(u,S-\log \frac{l}{r}) \\ \nonumber
    &+a \tilde{P}_{i}^Z(u,S-\log \frac{a}{c})+a\tilde{P}_{i}^Z(u,S-\log \frac{c}{a}) \\ \nonumber
    [a+c+u] \tilde{P}_i^Y(u,S) &= a \tilde{P}_{i}^X(u,S-\log \frac{a}{c})+c\tilde{P}_{i}^Z(u,S-\log \frac{c}{a}) \\ \nonumber
    [a+c+u] \tilde{P}_i^Z(u,S) &= c\tilde{P}_{i}^X(u,S-\log \frac{c}{a})+a\tilde{P}_{i}^Y(u,S-\log \frac{a}{c})
\end{eqnarray}
where we introduced the Laplace transforms $\tilde{P}_{i}^{\{X,Y,Z\}}(S,u)=\int_0^{\infty}dt e^{tu} P_{i}^{\{X,Y,Z\}}(S,t)$ and assumed initial condition on the $X$ state at $i=0$.
Upon exact decimation \cite{teza2020exact,teza2020thesis}, one can find the SCGF function of the original process to be the solution of the third-order polynomial equation
\begin{equation}\label{eqSM:second_loops_SCGF_3rd_order_poly}
    \varepsilon(\lambda)^3 + \alpha(\lambda) \varepsilon(\lambda)^2 +  \beta(\lambda) \varepsilon(\lambda) +  \gamma(\lambda) = 0  
\end{equation}
with coefficients
\begin{eqnarray}
    \alpha(\lambda) &&= 3 (a+c)+r+l -r e^{\lambda \log \frac{r}{l}} -l e^{\lambda \log \frac{l}{r}}\\ \nonumber
    \beta(\lambda) &&= 3(a^2+c^2+ac)+2(r+l)(a+c)
    -2(a+c))(r e^{\lambda \log \frac{r}{l}} +l e^{\lambda \log \frac{l}{r}})
    \\ \nonumber
    \gamma(\lambda) &&= a^3+c^3+(r+l)(a^2+c^2+ac)
    -a^3 e^{3\lambda \log \frac{a}{c}} -c^3 e^{3\lambda \log \frac{c}{a}}
    -(a^2+c^2+ac)(r e^{\lambda \log \frac{r}{l}} +l e^{\lambda \log \frac{l}{r}})
\end{eqnarray}
This provides direct access to all the cumulants of the EPR, yielding for the average
\begin{equation}
    \left< S/t \right>=(a-c) \log \frac{a}{c} + 3(r-l) \log \frac{r}{l} .
\end{equation}

Lumping together the secondary loops consisting into a triplet of $(X,Y,Z)$ states into a single effective state $\Xi$ provides us with a coarse-grained non-Markovian system evolving on the main loop (see sketch of Fig. \ref{figAP:secondary_loops_lumping} in the main text).
When the particle is hopping inside the secondary loop, it can produce entropy (in the case $a\neq c$) by performing a full revolution in either direction.
To not lose the contribution to the total EPR of these occurrences, we will therefore need to consider a full revolution as a "self-transition" from the state $\Xi$ to itself.
For convenience, we introduce the Laplace transforms of the exponential waiting time distributions of the original Markovian system $\tilde{\psi}_X(u)=\frac{r+l+a+c}{r+l+a+c+u}$ and  $\tilde{\psi}_Y(u)=\tilde{\psi}_Z(u)=\frac{a+c}{a+c+u}$. 
The lumped waiting time distribution therefore reads:
\begin{equation}
    \tilde{\Psi}(u)= \tilde{B}_0(u) \left(\frac{r+l}{r+l+a+c}\tilde{\psi}_X(u) +
    \frac{a^3+c^3}{(r+l+a+c)(a+c)^2} \tilde{\psi}_Y(u)\tilde{\psi}_Z(u) \tilde{B}_{XY}(u)\right)
\end{equation}
where $\tilde{B}_{XY}(u)=\sum_{k=0}^{\infty} \left(\frac{ac}{(a+c)^2}\tilde{\psi}_Y(u)\tilde{\psi}_Z(u)\right)^k$ accounts for the possibility of indefinitely bouncing back and forth between the $X$ and $Y$ states when performing a full revolution and $\tilde{B}_0(u)$ accounts for all the indefinite bounces that can be performed inside the loop without ever performing a full revolution.
Its explicit expression reads:
\begin{equation}
    \tilde{B}_0(u)=\left(1- \frac{ac}{(a+c)^2}\tilde{\psi}_Y(u)\tilde{\psi}_Z(u)\right)\sum_{k=0}^{\infty} \left(
    \frac{ac}{(a+c+r+l)(a+c)}\tilde{\psi}_X(u)\left(\tilde{\psi}_Y(u)+\tilde{\psi}_Z(u)\right)+
    \frac{ac}{(a+c)^2}\tilde{\psi}_Y(u)\tilde{\psi}_Z(u)
    \right)^k
\end{equation}
Writing everything explicitly in terms of the rates provides us ultimately with the following expression for the waiting time distribution:
\begin{equation}
    \tilde{\Psi}(u)=\frac{a^3+c^3+(l+r)\left(a^2+ac+c^2+2(a+c)u+u^2\right)}{a^3+c^3+(a^2+ac+c^2)(l+r)+\left(3(a^2+ac+c^2)+2(l+r)(a+c)\right)u+\left(3(a+c)+l+r\right)u^2+u^3}
\end{equation}
with the associated memory kernel
\begin{equation}
    \tilde{W}(u)=\frac{a^3+c^3+(l+r)\left(a^2+ac+c^2+2(a+c)u+u^2\right)}{3(a^2+ac+c^2)+3(a+c)s+s^2}
\end{equation}
Here, we can immediately distinguish the contributions to the memory kernel that are associated with a full revolution in the secondary loop (terms $\propto a^3$ and $\propto c^3$) and those coming from the transitions on the main loop (terms $\propto r$ and $\propto l$).
This allows us to write a generalized master equation for the probability $P^{\Xi}_i(t,S)$ of being in the lumped state$ \Xi$ at position $i$ at a given time $t$ system that properly accounts for the total produced entropy $S$:
{
\begin{eqnarray}\label{eqSM:second_loops_GME_laplace} \nonumber
  &&\left[a^3+c^3+(a^2+ac+c^2)(l+r)+\left(3(a^2+ac+c^2)+2(l+r)(a+c)\right)u+\left(3(a+c)+l+r\right)u^2+u^3\right](\tilde{P}^{\Xi}_i(u,S)-\delta_{i,0})= \\ \nonumber
&&=a^3\tilde{P}^{\Xi}_i(u,S-3\log \frac{a}{c})+
c^3\tilde{P}^{\Xi}_i(u,S-3\log \frac{c}{a})+ \\
&&+\left[a^2+ac+c^2+2(a+c)u+u^2\right](r\tilde{P}^{\Xi}_{i-1}(u,S-\log \frac{r}{l})+l\tilde{P}^{\Xi}_{i+1}(u,S-\log \frac{l}{r}))
\end{eqnarray}
}
We underline how summing over the entropy would provide us with the expected equation for $\tilde{P}^{\Xi}_i(u)=\sum_S \tilde{P}^{\Xi}_i(u,S)$ describing the spatial evolution on the main loop of the lumped system.

Inverse Laplace transforming Eq.~\ref{eqSM:second_loops_GME_laplace} allows us to obtain a third order differential equation for the cumulant generating function of the EPR of the lumped state $G'(\lambda,t)=\sum_{i,S}P^{\Xi}_ie^{\lambda S} (t,S)$.
In the approaching to the stationary state, the large deviation principle ensures that $G'(\lambda,t)\to e^{\varepsilon'(\lambda)t}$ with $\varepsilon'(\lambda)$ SCGF of the EPR of the lumped system.
Doing so, one recovers precisely the polynomial equation for the original SCGF (Eq.~\ref{eqSM:second_loops_SCGF_3rd_order_poly}), implying that the lumped description captures \textit{exactly} the entire EPR distribution of the original process.

This example shows on one hand how our lumping procedure allows to preserve all the entropy production also in an explicit case in which entire loops producing entropy are erased, while also showacasing a scenario in which the lumping and decimation procedure yield the same kind of coarse-grained non-Markovian dynamics.


\begin{thebibliography}{102}%
\makeatletter
\providecommand \@ifxundefined [1]{%
 \@ifx{#1\undefined}
}%
\providecommand \@ifnum [1]{%
 \ifnum #1\expandafter \@firstoftwo
 \else \expandafter \@secondoftwo
 \fi
}%
\providecommand \@ifx [1]{%
 \ifx #1\expandafter \@firstoftwo
 \else \expandafter \@secondoftwo
 \fi
}%
\providecommand \natexlab [1]{#1}%
\providecommand \enquote  [1]{``#1''}%
\providecommand \bibnamefont  [1]{#1}%
\providecommand \bibfnamefont [1]{#1}%
\providecommand \citenamefont [1]{#1}%
\providecommand \href@noop [0]{\@secondoftwo}%
\providecommand \href [0]{\begingroup \@sanitize@url \@href}%
\providecommand \@href[1]{\@@startlink{#1}\@@href}%
\providecommand \@@href[1]{\endgroup#1\@@endlink}%
\providecommand \@sanitize@url [0]{\catcode `\\12\catcode `\$12\catcode
  `\&12\catcode `\#12\catcode `\^12\catcode `\_12\catcode `\%12\relax}%
\providecommand \@@startlink[1]{}%
\providecommand \@@endlink[0]{}%
\providecommand \url  [0]{\begingroup\@sanitize@url \@url }%
\providecommand \@url [1]{\endgroup\@href {#1}{\urlprefix }}%
\providecommand \urlprefix  [0]{URL }%
\providecommand \Eprint [0]{\href }%
\providecommand \doibase [0]{https://doi.org/}%
\providecommand \selectlanguage [0]{\@gobble}%
\providecommand \bibinfo  [0]{\@secondoftwo}%
\providecommand \bibfield  [0]{\@secondoftwo}%
\providecommand \translation [1]{[#1]}%
\providecommand \BibitemOpen [0]{}%
\providecommand \bibitemStop [0]{}%
\providecommand \bibitemNoStop [0]{.\EOS\space}%
\providecommand \EOS [0]{\spacefactor3000\relax}%
\providecommand \BibitemShut  [1]{\csname bibitem#1\endcsname}%
\let\auto@bib@innerbib\@empty
\bibitem [{\citenamefont {Seifert}(2012)}]{seifert2012stochastic}%
  \BibitemOpen
  \bibfield  {author} {\bibinfo {author} {\bibfnamefont {U.}~\bibnamefont
  {Seifert}},\ }\bibfield  {title} {\bibinfo {title} {Stochastic
  thermodynamics, fluctuation theorems and molecular machines},\ }\href
  {https://doi.org/10.1088/0034-4885/75/12/126001} {\bibfield  {journal}
  {\bibinfo  {journal} {Reports on Progress in Physics}\ }\textbf {\bibinfo
  {volume} {75}},\ \bibinfo {pages} {126001} (\bibinfo {year}
  {2012})}\BibitemShut {NoStop}%
\bibitem [{\citenamefont {Speck}\ and\ \citenamefont
  {Seifert}(2006)}]{speck2006restoring}%
  \BibitemOpen
  \bibfield  {author} {\bibinfo {author} {\bibfnamefont {T.}~\bibnamefont
  {Speck}}\ and\ \bibinfo {author} {\bibfnamefont {U.}~\bibnamefont
  {Seifert}},\ }\bibfield  {title} {\bibinfo {title} {Restoring a
  fluctuation-dissipation theorem in a nonequilibrium steady state},\
  }\href@noop {} {\bibfield  {journal} {\bibinfo  {journal} {Europhysics
  Letters}\ }\textbf {\bibinfo {volume} {74}},\ \bibinfo {pages} {391}
  (\bibinfo {year} {2006})}\BibitemShut {NoStop}%
\bibitem [{\citenamefont {Prost}\ \emph {et~al.}(2009)\citenamefont {Prost},
  \citenamefont {Joanny},\ and\ \citenamefont
  {Parrondo}}]{prost2009generalized}%
  \BibitemOpen
  \bibfield  {author} {\bibinfo {author} {\bibfnamefont {J.}~\bibnamefont
  {Prost}}, \bibinfo {author} {\bibfnamefont {J.-F.}\ \bibnamefont {Joanny}},\
  and\ \bibinfo {author} {\bibfnamefont {J.~M.}\ \bibnamefont {Parrondo}},\
  }\bibfield  {title} {\bibinfo {title} {Generalized fluctuation-dissipation
  theorem for steady-state systems},\ }\href@noop {} {\bibfield  {journal}
  {\bibinfo  {journal} {Physical review letters}\ }\textbf {\bibinfo {volume}
  {103}},\ \bibinfo {pages} {090601} (\bibinfo {year} {2009})}\BibitemShut
  {NoStop}%
\bibitem [{\citenamefont {Baiesi}\ \emph {et~al.}(2009)\citenamefont {Baiesi},
  \citenamefont {Maes},\ and\ \citenamefont
  {Wynants}}]{baiesi2009fluctuations}%
  \BibitemOpen
  \bibfield  {author} {\bibinfo {author} {\bibfnamefont {M.}~\bibnamefont
  {Baiesi}}, \bibinfo {author} {\bibfnamefont {C.}~\bibnamefont {Maes}},\ and\
  \bibinfo {author} {\bibfnamefont {B.}~\bibnamefont {Wynants}},\ }\bibfield
  {title} {\bibinfo {title} {Fluctuations and response of nonequilibrium
  states},\ }\href@noop {} {\bibfield  {journal} {\bibinfo  {journal} {Physical
  review letters}\ }\textbf {\bibinfo {volume} {103}},\ \bibinfo {pages}
  {010602} (\bibinfo {year} {2009})}\BibitemShut {NoStop}%
\bibitem [{\citenamefont {Seifert}\ and\ \citenamefont
  {Speck}(2010)}]{seifert2010fluctuation}%
  \BibitemOpen
  \bibfield  {author} {\bibinfo {author} {\bibfnamefont {U.}~\bibnamefont
  {Seifert}}\ and\ \bibinfo {author} {\bibfnamefont {T.}~\bibnamefont
  {Speck}},\ }\bibfield  {title} {\bibinfo {title} {Fluctuation-dissipation
  theorem in nonequilibrium steady states},\ }\href@noop {} {\bibfield
  {journal} {\bibinfo  {journal} {Europhysics Letters}\ }\textbf {\bibinfo
  {volume} {89}},\ \bibinfo {pages} {10007} (\bibinfo {year}
  {2010})}\BibitemShut {NoStop}%
\bibitem [{\citenamefont {Chetrite}\ and\ \citenamefont
  {Gupta}(2011)}]{chetrite2011two}%
  \BibitemOpen
  \bibfield  {author} {\bibinfo {author} {\bibfnamefont {R.}~\bibnamefont
  {Chetrite}}\ and\ \bibinfo {author} {\bibfnamefont {S.}~\bibnamefont
  {Gupta}},\ }\bibfield  {title} {\bibinfo {title} {Two refreshing views of
  fluctuation theorems through kinematics elements and exponential
  martingale},\ }\href@noop {} {\bibfield  {journal} {\bibinfo  {journal}
  {Journal of Statistical Physics}\ }\textbf {\bibinfo {volume} {143}},\
  \bibinfo {pages} {543} (\bibinfo {year} {2011})}\BibitemShut {NoStop}%
\bibitem [{\citenamefont {Baiesi}\ and\ \citenamefont
  {Maes}(2013)}]{baiesi2013update}%
  \BibitemOpen
  \bibfield  {author} {\bibinfo {author} {\bibfnamefont {M.}~\bibnamefont
  {Baiesi}}\ and\ \bibinfo {author} {\bibfnamefont {C.}~\bibnamefont {Maes}},\
  }\bibfield  {title} {\bibinfo {title} {An update on the nonequilibrium linear
  response},\ }\href@noop {} {\bibfield  {journal} {\bibinfo  {journal} {New
  Journal of Physics}\ }\textbf {\bibinfo {volume} {15}},\ \bibinfo {pages}
  {013004} (\bibinfo {year} {2013})}\BibitemShut {NoStop}%
\bibitem [{\citenamefont {Maes}(2014)}]{maes2014second}%
  \BibitemOpen
  \bibfield  {author} {\bibinfo {author} {\bibfnamefont {C.}~\bibnamefont
  {Maes}},\ }\bibfield  {title} {\bibinfo {title} {On the second
  fluctuation--dissipation theorem for nonequilibrium baths},\ }\href@noop {}
  {\bibfield  {journal} {\bibinfo  {journal} {Journal of Statistical Physics}\
  }\textbf {\bibinfo {volume} {154}},\ \bibinfo {pages} {705} (\bibinfo {year}
  {2014})}\BibitemShut {NoStop}%
\bibitem [{\citenamefont {Kwon}\ \emph {et~al.}(2025)\citenamefont {Kwon},
  \citenamefont {Chun}, \citenamefont {Park},\ and\ \citenamefont
  {Lee}}]{kwon2025fluctuation}%
  \BibitemOpen
  \bibfield  {author} {\bibinfo {author} {\bibfnamefont {E.}~\bibnamefont
  {Kwon}}, \bibinfo {author} {\bibfnamefont {H.-M.}\ \bibnamefont {Chun}},
  \bibinfo {author} {\bibfnamefont {H.}~\bibnamefont {Park}},\ and\ \bibinfo
  {author} {\bibfnamefont {J.~S.}\ \bibnamefont {Lee}},\ }\bibfield  {title}
  {\bibinfo {title} {Fluctuation-response inequalities for kinetic and entropic
  perturbations},\ }\href@noop {} {\bibfield  {journal} {\bibinfo  {journal}
  {Physical Review Letters}\ }\textbf {\bibinfo {volume} {135}},\ \bibinfo
  {pages} {097101} (\bibinfo {year} {2025})}\BibitemShut {NoStop}%
\bibitem [{\citenamefont {Dechant}\ and\ \citenamefont
  {Sasa}(2020)}]{dechant2020fluctuation}%
  \BibitemOpen
  \bibfield  {author} {\bibinfo {author} {\bibfnamefont {A.}~\bibnamefont
  {Dechant}}\ and\ \bibinfo {author} {\bibfnamefont {S.-i.}\ \bibnamefont
  {Sasa}},\ }\bibfield  {title} {\bibinfo {title} {Fluctuation--response
  inequality out of equilibrium},\ }\href@noop {} {\bibfield  {journal}
  {\bibinfo  {journal} {Proceedings of the National Academy of Sciences}\
  }\textbf {\bibinfo {volume} {117}},\ \bibinfo {pages} {6430} (\bibinfo {year}
  {2020})}\BibitemShut {NoStop}%
\bibitem [{\citenamefont {Lebowitz}\ and\ \citenamefont
  {Spohn}(1999)}]{lebowitz1999gallavotti}%
  \BibitemOpen
  \bibfield  {author} {\bibinfo {author} {\bibfnamefont {J.~L.}\ \bibnamefont
  {Lebowitz}}\ and\ \bibinfo {author} {\bibfnamefont {H.}~\bibnamefont
  {Spohn}},\ }\bibfield  {title} {\bibinfo {title} {A gallavotti--cohen-type
  symmetry in the large deviation functional for stochastic dynamics},\
  }\href@noop {} {\bibfield  {journal} {\bibinfo  {journal} {Journal of
  Statistical Physics}\ }\textbf {\bibinfo {volume} {95}},\ \bibinfo {pages}
  {333} (\bibinfo {year} {1999})}\BibitemShut {NoStop}%
\bibitem [{\citenamefont {Jarzynski}(2006)}]{jarzynski2006rare}%
  \BibitemOpen
  \bibfield  {author} {\bibinfo {author} {\bibfnamefont {C.}~\bibnamefont
  {Jarzynski}},\ }\bibfield  {title} {\bibinfo {title} {Rare events and the
  convergence of exponentially averaged work values},\ }\href@noop {}
  {\bibfield  {journal} {\bibinfo  {journal} {Physical Review E}\ }\textbf
  {\bibinfo {volume} {73}},\ \bibinfo {pages} {046105} (\bibinfo {year}
  {2006})}\BibitemShut {NoStop}%
\bibitem [{\citenamefont {Jarzynski}(1997)}]{jarzynski1997nonequilibrium}%
  \BibitemOpen
  \bibfield  {author} {\bibinfo {author} {\bibfnamefont {C.}~\bibnamefont
  {Jarzynski}},\ }\bibfield  {title} {\bibinfo {title} {Nonequilibrium equality
  for free energy differences},\ }\href@noop {} {\bibfield  {journal} {\bibinfo
   {journal} {Physical Review Letters}\ }\textbf {\bibinfo {volume} {78}},\
  \bibinfo {pages} {2690} (\bibinfo {year} {1997})}\BibitemShut {NoStop}%
\bibitem [{\citenamefont {Searles}\ and\ \citenamefont
  {Evans}(1999)}]{searles1999fluctuation}%
  \BibitemOpen
  \bibfield  {author} {\bibinfo {author} {\bibfnamefont {D.~J.}\ \bibnamefont
  {Searles}}\ and\ \bibinfo {author} {\bibfnamefont {D.~J.}\ \bibnamefont
  {Evans}},\ }\bibfield  {title} {\bibinfo {title} {Fluctuation theorem for
  stochastic systems},\ }\href@noop {} {\bibfield  {journal} {\bibinfo
  {journal} {Physical Review E}\ }\textbf {\bibinfo {volume} {60}},\ \bibinfo
  {pages} {159} (\bibinfo {year} {1999})}\BibitemShut {NoStop}%
\bibitem [{\citenamefont {Kurchan}(1998)}]{kurchan1998fluctuation}%
  \BibitemOpen
  \bibfield  {author} {\bibinfo {author} {\bibfnamefont {J.}~\bibnamefont
  {Kurchan}},\ }\bibfield  {title} {\bibinfo {title} {Fluctuation theorem for
  stochastic dynamics},\ }\href@noop {} {\bibfield  {journal} {\bibinfo
  {journal} {Journal of Physics A: Mathematical and General}\ }\textbf
  {\bibinfo {volume} {31}},\ \bibinfo {pages} {3719} (\bibinfo {year}
  {1998})}\BibitemShut {NoStop}%
\bibitem [{\citenamefont {Crooks}(1999)}]{crooks1999entropy}%
  \BibitemOpen
  \bibfield  {author} {\bibinfo {author} {\bibfnamefont {G.~E.}\ \bibnamefont
  {Crooks}},\ }\bibfield  {title} {\bibinfo {title} {Entropy production
  fluctuation theorem and the nonequilibrium work relation for free energy
  differences},\ }\href@noop {} {\bibfield  {journal} {\bibinfo  {journal}
  {Physical Review E}\ }\textbf {\bibinfo {volume} {60}},\ \bibinfo {pages}
  {2721} (\bibinfo {year} {1999})}\BibitemShut {NoStop}%
\bibitem [{\citenamefont {Seifert}(2005)}]{seifert2005entropy}%
  \BibitemOpen
  \bibfield  {author} {\bibinfo {author} {\bibfnamefont {U.}~\bibnamefont
  {Seifert}},\ }\bibfield  {title} {\bibinfo {title} {Entropy production along
  a stochastic trajectory and an integral fluctuation theorem},\ }\href@noop {}
  {\bibfield  {journal} {\bibinfo  {journal} {Physical review letters}\
  }\textbf {\bibinfo {volume} {95}},\ \bibinfo {pages} {040602} (\bibinfo
  {year} {2005})}\BibitemShut {NoStop}%
\bibitem [{\citenamefont {Horowitz}\ and\ \citenamefont
  {Gingrich}(2020)}]{horowitz2020thermodynamic}%
  \BibitemOpen
  \bibfield  {author} {\bibinfo {author} {\bibfnamefont {J.~M.}\ \bibnamefont
  {Horowitz}}\ and\ \bibinfo {author} {\bibfnamefont {T.~R.}\ \bibnamefont
  {Gingrich}},\ }\bibfield  {title} {\bibinfo {title} {Thermodynamic
  uncertainty relations constrain non-equilibrium fluctuations},\ }\href@noop
  {} {\bibfield  {journal} {\bibinfo  {journal} {Nature Physics}\ }\textbf
  {\bibinfo {volume} {16}},\ \bibinfo {pages} {15} (\bibinfo {year}
  {2020})}\BibitemShut {NoStop}%
\bibitem [{\citenamefont {Gingrich}\ and\ \citenamefont
  {Horowitz}(2017)}]{gingrich2017fundamental}%
  \BibitemOpen
  \bibfield  {author} {\bibinfo {author} {\bibfnamefont {T.~R.}\ \bibnamefont
  {Gingrich}}\ and\ \bibinfo {author} {\bibfnamefont {J.~M.}\ \bibnamefont
  {Horowitz}},\ }\bibfield  {title} {\bibinfo {title} {Fundamental bounds on
  first passage time fluctuations for currents},\ }\href@noop {} {\bibfield
  {journal} {\bibinfo  {journal} {Physical review letters}\ }\textbf {\bibinfo
  {volume} {119}},\ \bibinfo {pages} {170601} (\bibinfo {year}
  {2017})}\BibitemShut {NoStop}%
\bibitem [{\citenamefont {Barato}\ and\ \citenamefont
  {Seifert}(2015)}]{barato2015thermodynamic}%
  \BibitemOpen
  \bibfield  {author} {\bibinfo {author} {\bibfnamefont {A.~C.}\ \bibnamefont
  {Barato}}\ and\ \bibinfo {author} {\bibfnamefont {U.}~\bibnamefont
  {Seifert}},\ }\bibfield  {title} {\bibinfo {title} {Thermodynamic uncertainty
  relation for biomolecular processes},\ }\href@noop {} {\bibfield  {journal}
  {\bibinfo  {journal} {Physical review letters}\ }\textbf {\bibinfo {volume}
  {114}},\ \bibinfo {pages} {158101} (\bibinfo {year} {2015})}\BibitemShut
  {NoStop}%
\bibitem [{\citenamefont {Gingrich}\ \emph {et~al.}(2016)\citenamefont
  {Gingrich}, \citenamefont {Horowitz}, \citenamefont {Perunov},\ and\
  \citenamefont {England}}]{gingrich2016dissipation}%
  \BibitemOpen
  \bibfield  {author} {\bibinfo {author} {\bibfnamefont {T.~R.}\ \bibnamefont
  {Gingrich}}, \bibinfo {author} {\bibfnamefont {J.~M.}\ \bibnamefont
  {Horowitz}}, \bibinfo {author} {\bibfnamefont {N.}~\bibnamefont {Perunov}},\
  and\ \bibinfo {author} {\bibfnamefont {J.~L.}\ \bibnamefont {England}},\
  }\bibfield  {title} {\bibinfo {title} {Dissipation bounds all steady-state
  current fluctuations},\ }\href@noop {} {\bibfield  {journal} {\bibinfo
  {journal} {Physical review letters}\ }\textbf {\bibinfo {volume} {116}},\
  \bibinfo {pages} {120601} (\bibinfo {year} {2016})}\BibitemShut {NoStop}%
\bibitem [{\citenamefont {Polettini}\ \emph {et~al.}(2016)\citenamefont
  {Polettini}, \citenamefont {Lazarescu},\ and\ \citenamefont
  {Esposito}}]{polettini2016tightening}%
  \BibitemOpen
  \bibfield  {author} {\bibinfo {author} {\bibfnamefont {M.}~\bibnamefont
  {Polettini}}, \bibinfo {author} {\bibfnamefont {A.}~\bibnamefont
  {Lazarescu}},\ and\ \bibinfo {author} {\bibfnamefont {M.}~\bibnamefont
  {Esposito}},\ }\bibfield  {title} {\bibinfo {title} {Tightening the
  uncertainty principle for stochastic currents},\ }\href@noop {} {\bibfield
  {journal} {\bibinfo  {journal} {Physical Review E}\ }\textbf {\bibinfo
  {volume} {94}},\ \bibinfo {pages} {052104} (\bibinfo {year}
  {2016})}\BibitemShut {NoStop}%
\bibitem [{\citenamefont {Dechant}\ \emph {et~al.}(2025)\citenamefont
  {Dechant}, \citenamefont {Saito} \emph {et~al.}}]{dechant2025inverse}%
  \BibitemOpen
  \bibfield  {author} {\bibinfo {author} {\bibfnamefont {A.}~\bibnamefont
  {Dechant}}, \bibinfo {author} {\bibfnamefont {K.}~\bibnamefont {Saito}},
  \emph {et~al.},\ }\bibfield  {title} {\bibinfo {title} {Inverse thermodynamic
  uncertainty relation and entropy production},\ }\href@noop {} {\bibfield
  {journal} {\bibinfo  {journal} {Physical Review Letters}\ }\textbf {\bibinfo
  {volume} {135}},\ \bibinfo {pages} {237104} (\bibinfo {year}
  {2025})}\BibitemShut {NoStop}%
\bibitem [{\citenamefont {Raghu}\ and\ \citenamefont
  {Neri}(2025)}]{raghu2025thermodynamic}%
  \BibitemOpen
  \bibfield  {author} {\bibinfo {author} {\bibfnamefont {A.}~\bibnamefont
  {Raghu}}\ and\ \bibinfo {author} {\bibfnamefont {I.}~\bibnamefont {Neri}},\
  }\bibfield  {title} {\bibinfo {title} {Thermodynamic bounds and symmetries in
  first-passage problems of fluctuating currents},\ }\href@noop {} {\bibfield
  {journal} {\bibinfo  {journal} {New Journal of Physics}\ } (\bibinfo {year}
  {2025})}\BibitemShut {NoStop}%
\bibitem [{\citenamefont {Di~Terlizzi}\ and\ \citenamefont
  {Baiesi}(2018)}]{di2018kinetic}%
  \BibitemOpen
  \bibfield  {author} {\bibinfo {author} {\bibfnamefont {I.}~\bibnamefont
  {Di~Terlizzi}}\ and\ \bibinfo {author} {\bibfnamefont {M.}~\bibnamefont
  {Baiesi}},\ }\bibfield  {title} {\bibinfo {title} {Kinetic uncertainty
  relation},\ }\href@noop {} {\bibfield  {journal} {\bibinfo  {journal}
  {Journal of Physics A: Mathematical and Theoretical}\ }\textbf {\bibinfo
  {volume} {52}},\ \bibinfo {pages} {02LT03} (\bibinfo {year}
  {2018})}\BibitemShut {NoStop}%
\bibitem [{\citenamefont {Van~Vu}\ \emph {et~al.}(2022)\citenamefont {Van~Vu},
  \citenamefont {Hasegawa} \emph {et~al.}}]{van2022unified}%
  \BibitemOpen
  \bibfield  {author} {\bibinfo {author} {\bibfnamefont {T.}~\bibnamefont
  {Van~Vu}}, \bibinfo {author} {\bibfnamefont {Y.}~\bibnamefont {Hasegawa}},
  \emph {et~al.},\ }\bibfield  {title} {\bibinfo {title} {Unified
  thermodynamic--kinetic uncertainty relation},\ }\href@noop {} {\bibfield
  {journal} {\bibinfo  {journal} {Journal of Physics A: Mathematical and
  Theoretical}\ }\textbf {\bibinfo {volume} {55}},\ \bibinfo {pages} {405004}
  (\bibinfo {year} {2022})}\BibitemShut {NoStop}%
\bibitem [{\citenamefont {Palmqvist}\ \emph {et~al.}(2025)\citenamefont
  {Palmqvist}, \citenamefont {Tesser},\ and\ \citenamefont
  {Splettstoesser}}]{palmqvist2025kinetic}%
  \BibitemOpen
  \bibfield  {author} {\bibinfo {author} {\bibfnamefont {D.}~\bibnamefont
  {Palmqvist}}, \bibinfo {author} {\bibfnamefont {L.}~\bibnamefont {Tesser}},\
  and\ \bibinfo {author} {\bibfnamefont {J.}~\bibnamefont {Splettstoesser}},\
  }\bibfield  {title} {\bibinfo {title} {Kinetic uncertainty relations for
  quantum transport},\ }\href@noop {} {\bibfield  {journal} {\bibinfo
  {journal} {Physical Review Letters}\ }\textbf {\bibinfo {volume} {135}},\
  \bibinfo {pages} {166302} (\bibinfo {year} {2025})}\BibitemShut {NoStop}%
\bibitem [{\citenamefont {Hiura}\ and\ \citenamefont
  {Sasa}(2021)}]{hiura2021kinetic}%
  \BibitemOpen
  \bibfield  {author} {\bibinfo {author} {\bibfnamefont {K.}~\bibnamefont
  {Hiura}}\ and\ \bibinfo {author} {\bibfnamefont {S.-i.}\ \bibnamefont
  {Sasa}},\ }\bibfield  {title} {\bibinfo {title} {Kinetic uncertainty relation
  on first-passage time for accumulated current},\ }\href@noop {} {\bibfield
  {journal} {\bibinfo  {journal} {Physical Review E}\ }\textbf {\bibinfo
  {volume} {103}},\ \bibinfo {pages} {L050103} (\bibinfo {year}
  {2021})}\BibitemShut {NoStop}%
\bibitem [{\citenamefont {Yan}\ \emph {et~al.}(2019)\citenamefont {Yan},
  \citenamefont {Hilfinger}, \citenamefont {Vinnicombe},\ and\ \citenamefont
  {Paulsson}}]{yan2019kinetic}%
  \BibitemOpen
  \bibfield  {author} {\bibinfo {author} {\bibfnamefont {J.}~\bibnamefont
  {Yan}}, \bibinfo {author} {\bibfnamefont {A.}~\bibnamefont {Hilfinger}},
  \bibinfo {author} {\bibfnamefont {G.}~\bibnamefont {Vinnicombe}},\ and\
  \bibinfo {author} {\bibfnamefont {J.}~\bibnamefont {Paulsson}},\ }\bibfield
  {title} {\bibinfo {title} {Kinetic uncertainty relations for the control of
  stochastic reaction networks},\ }\href@noop {} {\bibfield  {journal}
  {\bibinfo  {journal} {Physical review letters}\ }\textbf {\bibinfo {volume}
  {123}},\ \bibinfo {pages} {108101} (\bibinfo {year} {2019})}\BibitemShut
  {NoStop}%
\bibitem [{\citenamefont {Seifert}(2025)}]{seifert2025universal}%
  \BibitemOpen
  \bibfield  {author} {\bibinfo {author} {\bibfnamefont {U.}~\bibnamefont
  {Seifert}},\ }\bibfield  {title} {\bibinfo {title} {Universal bounds on
  entropy production from fluctuating coarse-grained trajectories},\ }\bibfield
   {journal} {\bibinfo  {journal} {arXiv preprint arXiv:2512.07772}\ }\href
  {https://doi.org/10.48550/arXiv.2512.07772} {10.48550/arXiv.2512.07772}
  (\bibinfo {year} {2025})\BibitemShut {NoStop}%
\bibitem [{\citenamefont {Mart{\'\i}nez}\ \emph {et~al.}(2019)\citenamefont
  {Mart{\'\i}nez}, \citenamefont {Bisker}, \citenamefont {Horowitz},\ and\
  \citenamefont {Parrondo}}]{martinez2019inferring}%
  \BibitemOpen
  \bibfield  {author} {\bibinfo {author} {\bibfnamefont {I.~A.}\ \bibnamefont
  {Mart{\'\i}nez}}, \bibinfo {author} {\bibfnamefont {G.}~\bibnamefont
  {Bisker}}, \bibinfo {author} {\bibfnamefont {J.~M.}\ \bibnamefont
  {Horowitz}},\ and\ \bibinfo {author} {\bibfnamefont {J.~M.}\ \bibnamefont
  {Parrondo}},\ }\bibfield  {title} {\bibinfo {title} {Inferring broken
  detailed balance in the absence of observable currents},\ }\href@noop {}
  {\bibfield  {journal} {\bibinfo  {journal} {Nature communications}\ }\textbf
  {\bibinfo {volume} {10}},\ \bibinfo {pages} {3542} (\bibinfo {year}
  {2019})}\BibitemShut {NoStop}%
\bibitem [{\citenamefont {van~der Meer}\ \emph {et~al.}(2023)\citenamefont
  {van~der Meer}, \citenamefont {Deg{\"u}nther},\ and\ \citenamefont
  {Seifert}}]{van2023time}%
  \BibitemOpen
  \bibfield  {author} {\bibinfo {author} {\bibfnamefont {J.}~\bibnamefont
  {van~der Meer}}, \bibinfo {author} {\bibfnamefont {J.}~\bibnamefont
  {Deg{\"u}nther}},\ and\ \bibinfo {author} {\bibfnamefont {U.}~\bibnamefont
  {Seifert}},\ }\bibfield  {title} {\bibinfo {title} {Time-resolved statistics
  of snippets as general framework for model-free entropy estimators},\
  }\href@noop {} {\bibfield  {journal} {\bibinfo  {journal} {Physical Review
  Letters}\ }\textbf {\bibinfo {volume} {130}},\ \bibinfo {pages} {257101}
  (\bibinfo {year} {2023})}\BibitemShut {NoStop}%
\bibitem [{\citenamefont {Harunari}\ \emph {et~al.}(2022)\citenamefont
  {Harunari}, \citenamefont {Dutta}, \citenamefont {Polettini},\ and\
  \citenamefont {Rold{\'a}n}}]{harunari2022learn}%
  \BibitemOpen
  \bibfield  {author} {\bibinfo {author} {\bibfnamefont {P.~E.}\ \bibnamefont
  {Harunari}}, \bibinfo {author} {\bibfnamefont {A.}~\bibnamefont {Dutta}},
  \bibinfo {author} {\bibfnamefont {M.}~\bibnamefont {Polettini}},\ and\
  \bibinfo {author} {\bibfnamefont {{\'E}.}~\bibnamefont {Rold{\'a}n}},\
  }\bibfield  {title} {\bibinfo {title} {What to learn from a few visible
  transitions’ statistics?},\ }\href@noop {} {\bibfield  {journal} {\bibinfo
  {journal} {Physical Review X}\ }\textbf {\bibinfo {volume} {12}},\ \bibinfo
  {pages} {041026} (\bibinfo {year} {2022})}\BibitemShut {NoStop}%
\bibitem [{\citenamefont {Ehrich}(2021)}]{ehrich2021tightest}%
  \BibitemOpen
  \bibfield  {author} {\bibinfo {author} {\bibfnamefont {J.}~\bibnamefont
  {Ehrich}},\ }\bibfield  {title} {\bibinfo {title} {Tightest bound on hidden
  entropy production from partially observed dynamics},\ }\href@noop {}
  {\bibfield  {journal} {\bibinfo  {journal} {Journal of Statistical Mechanics:
  Theory and Experiment}\ }\textbf {\bibinfo {volume} {2021}},\ \bibinfo
  {pages} {083214} (\bibinfo {year} {2021})}\BibitemShut {NoStop}%
\bibitem [{\citenamefont {Ghosal}\ and\ \citenamefont
  {Bisker}(2022)}]{ghosal2022inferring}%
  \BibitemOpen
  \bibfield  {author} {\bibinfo {author} {\bibfnamefont {A.}~\bibnamefont
  {Ghosal}}\ and\ \bibinfo {author} {\bibfnamefont {G.}~\bibnamefont
  {Bisker}},\ }\bibfield  {title} {\bibinfo {title} {Inferring entropy
  production rate from partially observed langevin dynamics under
  coarse-graining},\ }\href@noop {} {\bibfield  {journal} {\bibinfo  {journal}
  {Physical Chemistry Chemical Physics}\ }\textbf {\bibinfo {volume} {24}},\
  \bibinfo {pages} {24021} (\bibinfo {year} {2022})}\BibitemShut {NoStop}%
\bibitem [{\citenamefont {Pietzonka}\ and\ \citenamefont
  {Coghi}(2024)}]{pietzonka2024thermodynamic}%
  \BibitemOpen
  \bibfield  {author} {\bibinfo {author} {\bibfnamefont {P.}~\bibnamefont
  {Pietzonka}}\ and\ \bibinfo {author} {\bibfnamefont {F.}~\bibnamefont
  {Coghi}},\ }\bibfield  {title} {\bibinfo {title} {Thermodynamic cost for
  precision of general counting observables},\ }\href@noop {} {\bibfield
  {journal} {\bibinfo  {journal} {Physical Review E}\ }\textbf {\bibinfo
  {volume} {109}},\ \bibinfo {pages} {064128} (\bibinfo {year}
  {2024})}\BibitemShut {NoStop}%
\bibitem [{\citenamefont {Baiesi}\ \emph {et~al.}(2024)\citenamefont {Baiesi},
  \citenamefont {Nishiyama},\ and\ \citenamefont
  {Falasco}}]{baiesi2024effective}%
  \BibitemOpen
  \bibfield  {author} {\bibinfo {author} {\bibfnamefont {M.}~\bibnamefont
  {Baiesi}}, \bibinfo {author} {\bibfnamefont {T.}~\bibnamefont {Nishiyama}},\
  and\ \bibinfo {author} {\bibfnamefont {G.}~\bibnamefont {Falasco}},\
  }\bibfield  {title} {\bibinfo {title} {Effective estimation of entropy
  production with lacking data},\ }\href@noop {} {\bibfield  {journal}
  {\bibinfo  {journal} {Communications Physics}\ }\textbf {\bibinfo {volume}
  {7}},\ \bibinfo {pages} {264} (\bibinfo {year} {2024})}\BibitemShut {NoStop}%
\bibitem [{\citenamefont {Skinner}\ and\ \citenamefont
  {Dunkel}(2021)}]{skinner2021estimating}%
  \BibitemOpen
  \bibfield  {author} {\bibinfo {author} {\bibfnamefont {D.~J.}\ \bibnamefont
  {Skinner}}\ and\ \bibinfo {author} {\bibfnamefont {J.}~\bibnamefont
  {Dunkel}},\ }\bibfield  {title} {\bibinfo {title} {Estimating entropy
  production from waiting time distributions},\ }\href@noop {} {\bibfield
  {journal} {\bibinfo  {journal} {Physical review letters}\ }\textbf {\bibinfo
  {volume} {127}},\ \bibinfo {pages} {198101} (\bibinfo {year}
  {2021})}\BibitemShut {NoStop}%
\bibitem [{\citenamefont {Blom}\ \emph {et~al.}(2024)\citenamefont {Blom},
  \citenamefont {Song}, \citenamefont {Vouga}, \citenamefont {Godec},\ and\
  \citenamefont {Makarov}}]{blom2024milestoning}%
  \BibitemOpen
  \bibfield  {author} {\bibinfo {author} {\bibfnamefont {K.}~\bibnamefont
  {Blom}}, \bibinfo {author} {\bibfnamefont {K.}~\bibnamefont {Song}}, \bibinfo
  {author} {\bibfnamefont {E.}~\bibnamefont {Vouga}}, \bibinfo {author}
  {\bibfnamefont {A.}~\bibnamefont {Godec}},\ and\ \bibinfo {author}
  {\bibfnamefont {D.~E.}\ \bibnamefont {Makarov}},\ }\bibfield  {title}
  {\bibinfo {title} {Milestoning estimators of dissipation in systems observed
  at a coarse resolution},\ }\href@noop {} {\bibfield  {journal} {\bibinfo
  {journal} {Proceedings of the National Academy of Sciences}\ }\textbf
  {\bibinfo {volume} {121}},\ \bibinfo {pages} {e2318333121} (\bibinfo {year}
  {2024})}\BibitemShut {NoStop}%
\bibitem [{\citenamefont {Di~Terlizzi}(2025)}]{diterlizzi2025forcefree}%
  \BibitemOpen
  \bibfield  {author} {\bibinfo {author} {\bibfnamefont {I.}~\bibnamefont
  {Di~Terlizzi}},\ }\bibfield  {title} {\bibinfo {title} {Force-free kinetic
  inference of entropy production},\ }\href {https://doi.org/10.1103/fsph-437v}
  {\bibfield  {journal} {\bibinfo  {journal} {Phys. Rev. Lett.}\ }\textbf
  {\bibinfo {volume} {135}},\ \bibinfo {pages} {237101} (\bibinfo {year}
  {2025})}\BibitemShut {NoStop}%
\bibitem [{\citenamefont {Pekola}(2015)}]{pekola2015towards}%
  \BibitemOpen
  \bibfield  {author} {\bibinfo {author} {\bibfnamefont {J.~P.}\ \bibnamefont
  {Pekola}},\ }\bibfield  {title} {\bibinfo {title} {Towards quantum
  thermodynamics in electronic circuits},\ }\href@noop {} {\bibfield  {journal}
  {\bibinfo  {journal} {Nature physics}\ }\textbf {\bibinfo {volume} {11}},\
  \bibinfo {pages} {118} (\bibinfo {year} {2015})}\BibitemShut {NoStop}%
\bibitem [{\citenamefont {Ciliberto}(2017)}]{ciliberto2017}%
  \BibitemOpen
  \bibfield  {author} {\bibinfo {author} {\bibfnamefont {S.}~\bibnamefont
  {Ciliberto}},\ }\bibfield  {title} {\bibinfo {title} {Experiments in
  stochastic thermodynamics: Short history and perspectives},\ }\href@noop {}
  {\bibfield  {journal} {\bibinfo  {journal} {Physical Review X}\ }\textbf
  {\bibinfo {volume} {7}},\ \bibinfo {pages} {021051} (\bibinfo {year}
  {2017})}\BibitemShut {NoStop}%
\bibitem [{\citenamefont {Alemany}\ \emph {et~al.}(2015)\citenamefont
  {Alemany}, \citenamefont {Ribezzi-Crivellari},\ and\ \citenamefont
  {Ritort}}]{alemany2015}%
  \BibitemOpen
  \bibfield  {author} {\bibinfo {author} {\bibfnamefont {A.}~\bibnamefont
  {Alemany}}, \bibinfo {author} {\bibfnamefont {M.}~\bibnamefont
  {Ribezzi-Crivellari}},\ and\ \bibinfo {author} {\bibfnamefont
  {F.}~\bibnamefont {Ritort}},\ }\bibfield  {title} {\bibinfo {title} {From
  free energy measurements to thermodynamic inference in nonequilibrium small
  systems},\ }\href@noop {} {\bibfield  {journal} {\bibinfo  {journal} {New
  Journal of Physics}\ }\textbf {\bibinfo {volume} {17}},\ \bibinfo {pages}
  {075009} (\bibinfo {year} {2015})}\BibitemShut {NoStop}%
\bibitem [{\citenamefont {Gnesotto}\ \emph {et~al.}(2018)\citenamefont
  {Gnesotto}, \citenamefont {Mura}, \citenamefont {Gladrow},\ and\
  \citenamefont {Broedersz}}]{gnesotto2018}%
  \BibitemOpen
  \bibfield  {author} {\bibinfo {author} {\bibfnamefont {F.}~\bibnamefont
  {Gnesotto}}, \bibinfo {author} {\bibfnamefont {F.}~\bibnamefont {Mura}},
  \bibinfo {author} {\bibfnamefont {J.}~\bibnamefont {Gladrow}},\ and\ \bibinfo
  {author} {\bibfnamefont {C.}~\bibnamefont {Broedersz}},\ }\bibfield  {title}
  {\bibinfo {title} {Reports on progress in physics broken detailed balance and
  non-equilibrium dynamics in living systems: a review broken detailed balance
  and non-equilibrium dynamics in living systems: a review review},\
  }\href@noop {} {\bibfield  {journal} {\bibinfo  {journal} {Rep. Prog. Phys}\
  }\textbf {\bibinfo {volume} {81}},\ \bibinfo {pages} {32} (\bibinfo {year}
  {2018})}\BibitemShut {NoStop}%
\bibitem [{\citenamefont {Lau}\ \emph {et~al.}(2007)\citenamefont {Lau},
  \citenamefont {Lacoste},\ and\ \citenamefont
  {Mallick}}]{mallick2007nonequilibrium}%
  \BibitemOpen
  \bibfield  {author} {\bibinfo {author} {\bibfnamefont {A.~W.~C.}\
  \bibnamefont {Lau}}, \bibinfo {author} {\bibfnamefont {D.}~\bibnamefont
  {Lacoste}},\ and\ \bibinfo {author} {\bibfnamefont {K.}~\bibnamefont
  {Mallick}},\ }\bibfield  {title} {\bibinfo {title} {Nonequilibrium
  fluctuations and mechanochemical couplings of a molecular motor},\ }\href
  {https://doi.org/10.1103/PhysRevLett.99.158102} {\bibfield  {journal}
  {\bibinfo  {journal} {Phys. Rev. Lett.}\ }\textbf {\bibinfo {volume} {99}},\
  \bibinfo {pages} {158102} (\bibinfo {year} {2007})}\BibitemShut {NoStop}%
\bibitem [{\citenamefont {Teza}\ and\ \citenamefont
  {Stella}(2020)}]{teza2020exact}%
  \BibitemOpen
  \bibfield  {author} {\bibinfo {author} {\bibfnamefont {G.}~\bibnamefont
  {Teza}}\ and\ \bibinfo {author} {\bibfnamefont {A.~L.}\ \bibnamefont
  {Stella}},\ }\bibfield  {title} {\bibinfo {title} {Exact coarse graining
  preserves entropy production out of equilibrium},\ }\href@noop {} {\bibfield
  {journal} {\bibinfo  {journal} {Physical Review Letters}\ }\textbf {\bibinfo
  {volume} {125}},\ \bibinfo {pages} {110601} (\bibinfo {year}
  {2020})}\BibitemShut {NoStop}%
\bibitem [{\citenamefont {Dieball}\ and\ \citenamefont
  {Godec}(2025)}]{dieball2025perspective}%
  \BibitemOpen
  \bibfield  {author} {\bibinfo {author} {\bibfnamefont {C.}~\bibnamefont
  {Dieball}}\ and\ \bibinfo {author} {\bibfnamefont {A.}~\bibnamefont
  {Godec}},\ }\bibfield  {title} {\bibinfo {title} {Perspective: Time
  irreversibility in systems observed at coarse resolution},\ }\href@noop {}
  {\bibfield  {journal} {\bibinfo  {journal} {The Journal of Chemical Physics}\
  }\textbf {\bibinfo {volume} {162}} (\bibinfo {year} {2025})}\BibitemShut
  {NoStop}%
\bibitem [{\citenamefont {Geiger}\ and\ \citenamefont
  {Temmel}(2014)}]{geiger2014lumpings}%
  \BibitemOpen
  \bibfield  {author} {\bibinfo {author} {\bibfnamefont {B.~C.}\ \bibnamefont
  {Geiger}}\ and\ \bibinfo {author} {\bibfnamefont {C.}~\bibnamefont
  {Temmel}},\ }\bibfield  {title} {\bibinfo {title} {Lumpings of markov chains,
  entropy rate preservation, and higher-order lumpability},\ }\href@noop {}
  {\bibfield  {journal} {\bibinfo  {journal} {Journal of Applied Probability}\
  }\textbf {\bibinfo {volume} {51}},\ \bibinfo {pages} {1114} (\bibinfo {year}
  {2014})}\BibitemShut {NoStop}%
\bibitem [{\citenamefont {Hoffmann}\ and\ \citenamefont
  {Salamon}(2009)}]{hoffmann2009bounding}%
  \BibitemOpen
  \bibfield  {author} {\bibinfo {author} {\bibfnamefont {K.~H.}\ \bibnamefont
  {Hoffmann}}\ and\ \bibinfo {author} {\bibfnamefont {P.}~\bibnamefont
  {Salamon}},\ }\bibfield  {title} {\bibinfo {title} {Bounding the lumping
  error in markov chain dynamics},\ }\href@noop {} {\bibfield  {journal}
  {\bibinfo  {journal} {Applied mathematics letters}\ }\textbf {\bibinfo
  {volume} {22}},\ \bibinfo {pages} {1471} (\bibinfo {year}
  {2009})}\BibitemShut {NoStop}%
\bibitem [{\citenamefont {Igoshin}\ \emph {et~al.}(2025)\citenamefont
  {Igoshin}, \citenamefont {Kolomeisky},\ and\ \citenamefont
  {Makarov}}]{igoshin2025uncovering}%
  \BibitemOpen
  \bibfield  {author} {\bibinfo {author} {\bibfnamefont {O.~A.}\ \bibnamefont
  {Igoshin}}, \bibinfo {author} {\bibfnamefont {A.~B.}\ \bibnamefont
  {Kolomeisky}},\ and\ \bibinfo {author} {\bibfnamefont {D.~E.}\ \bibnamefont
  {Makarov}},\ }\bibfield  {title} {\bibinfo {title} {Uncovering dissipation
  from coarse observables: A case study of a random walk with unobserved
  internal states},\ }\href@noop {} {\bibfield  {journal} {\bibinfo  {journal}
  {The Journal of Chemical Physics}\ }\textbf {\bibinfo {volume} {162}}
  (\bibinfo {year} {2025})}\BibitemShut {NoStop}%
\bibitem [{\citenamefont {Esposito}(2012)}]{esposito2012stochastic}%
  \BibitemOpen
  \bibfield  {author} {\bibinfo {author} {\bibfnamefont {M.}~\bibnamefont
  {Esposito}},\ }\bibfield  {title} {\bibinfo {title} {Stochastic
  thermodynamics under coarse graining},\ }\href
  {https://doi.org/10.1103/PhysRevE.85.041125} {\bibfield  {journal} {\bibinfo
  {journal} {Phys. Rev. E}\ }\textbf {\bibinfo {volume} {85}},\ \bibinfo
  {pages} {041125} (\bibinfo {year} {2012})}\BibitemShut {NoStop}%
\bibitem [{\citenamefont {Rahav}\ and\ \citenamefont
  {Jarzynski}(2007)}]{rahav2007fluctuation}%
  \BibitemOpen
  \bibfield  {author} {\bibinfo {author} {\bibfnamefont {S.}~\bibnamefont
  {Rahav}}\ and\ \bibinfo {author} {\bibfnamefont {C.}~\bibnamefont
  {Jarzynski}},\ }\bibfield  {title} {\bibinfo {title} {Fluctuation relations
  and coarse-graining},\ }\href
  {https://doi.org/10.1088/1742-5468/2007/09/P09012} {\bibfield  {journal}
  {\bibinfo  {journal} {Journal of Statistical Mechanics: Theory and
  Experiment}\ }\textbf {\bibinfo {volume} {2007}},\ \bibinfo {pages} {P09012}
  (\bibinfo {year} {2007})}\BibitemShut {NoStop}%
\bibitem [{\citenamefont {Amann}\ \emph {et~al.}(2010)\citenamefont {Amann},
  \citenamefont {Schmiedl},\ and\ \citenamefont
  {Seifert}}]{amann2010communications}%
  \BibitemOpen
  \bibfield  {author} {\bibinfo {author} {\bibfnamefont {C.~P.}\ \bibnamefont
  {Amann}}, \bibinfo {author} {\bibfnamefont {T.}~\bibnamefont {Schmiedl}},\
  and\ \bibinfo {author} {\bibfnamefont {U.}~\bibnamefont {Seifert}},\
  }\bibfield  {title} {\bibinfo {title} {{Communications: Can one identify
  nonequilibrium in a three-state system by analyzing two-state
  trajectories?}},\ }\href {https://doi.org/10.1063/1.3294567} {\bibfield
  {journal} {\bibinfo  {journal} {The Journal of Chemical Physics}\ }\textbf
  {\bibinfo {volume} {132}},\ \bibinfo {pages} {041102} (\bibinfo {year}
  {2010})},\ \Eprint
  {https://arxiv.org/abs/https://pubs.aip.org/aip/jcp/article-pdf/doi/10.1063/1.3294567/16029179/041102\_1\_online.pdf}
  {https://pubs.aip.org/aip/jcp/article-pdf/doi/10.1063/1.3294567/16029179/041102\_1\_online.pdf}
  \BibitemShut {NoStop}%
\bibitem [{\citenamefont {Nicolis}(2011)}]{nicolis2011transformation}%
  \BibitemOpen
  \bibfield  {author} {\bibinfo {author} {\bibfnamefont {G.}~\bibnamefont
  {Nicolis}},\ }\bibfield  {title} {\bibinfo {title} {Transformation properties
  of entropy production},\ }\href@noop {} {\bibfield  {journal} {\bibinfo
  {journal} {Physical Review E—Statistical, Nonlinear, and Soft Matter
  Physics}\ }\textbf {\bibinfo {volume} {83}},\ \bibinfo {pages} {011112}
  (\bibinfo {year} {2011})}\BibitemShut {NoStop}%
\bibitem [{\citenamefont {Hartich}\ and\ \citenamefont
  {Godec}(2023)}]{hartich2023violation}%
  \BibitemOpen
  \bibfield  {author} {\bibinfo {author} {\bibfnamefont {D.}~\bibnamefont
  {Hartich}}\ and\ \bibinfo {author} {\bibfnamefont {A.}~\bibnamefont
  {Godec}},\ }\bibfield  {title} {\bibinfo {title} {Violation of local detailed
  balance upon lumping despite a clear timescale separation},\ }\href@noop {}
  {\bibfield  {journal} {\bibinfo  {journal} {Physical Review Research}\
  }\textbf {\bibinfo {volume} {5}},\ \bibinfo {pages} {L032017} (\bibinfo
  {year} {2023})}\BibitemShut {NoStop}%
\bibitem [{\citenamefont {Tabanera-Bravo}\ and\ \citenamefont
  {Godec}(2025)}]{tabanera2025purely}%
  \BibitemOpen
  \bibfield  {author} {\bibinfo {author} {\bibfnamefont {J.}~\bibnamefont
  {Tabanera-Bravo}}\ and\ \bibinfo {author} {\bibfnamefont {A.}~\bibnamefont
  {Godec}},\ }\bibfield  {title} {\bibinfo {title} {Purely quantum memory in
  closed systems observed via imperfect measurements},\ }\href@noop {}
  {\bibfield  {journal} {\bibinfo  {journal} {arXiv preprint arXiv:2506.13689}\
  } (\bibinfo {year} {2025})}\BibitemShut {NoStop}%
\bibitem [{\citenamefont {Hughes}(1995)}]{hughes1995random}%
  \BibitemOpen
  \bibfield  {author} {\bibinfo {author} {\bibfnamefont {B.~D.}\ \bibnamefont
  {Hughes}},\ }\href@noop {} {\emph {\bibinfo {title} {Random walks and random
  environments: random walks}}},\ Vol.~\bibinfo {volume} {1}\ (\bibinfo
  {publisher} {Oxford University Press},\ \bibinfo {year} {1995})\BibitemShut
  {NoStop}%
\bibitem [{\citenamefont {Esposito}\ and\ \citenamefont
  {Lindenberg}(2008)}]{esposito2008continuous}%
  \BibitemOpen
  \bibfield  {author} {\bibinfo {author} {\bibfnamefont {M.}~\bibnamefont
  {Esposito}}\ and\ \bibinfo {author} {\bibfnamefont {K.}~\bibnamefont
  {Lindenberg}},\ }\bibfield  {title} {\bibinfo {title} {Continuous-time random
  walk for open systems: Fluctuation theorems and counting statistics},\ }\href
  {https://doi.org/10.1103/PhysRevE.77.051119} {\bibfield  {journal} {\bibinfo
  {journal} {Phys. Rev. E}\ }\textbf {\bibinfo {volume} {77}},\ \bibinfo
  {pages} {051119} (\bibinfo {year} {2008})}\BibitemShut {NoStop}%
\bibitem [{\citenamefont {Andrieux}\ and\ \citenamefont
  {Gaspard}(2008)}]{andrieux2008thefluctuation}%
  \BibitemOpen
  \bibfield  {author} {\bibinfo {author} {\bibfnamefont {D.}~\bibnamefont
  {Andrieux}}\ and\ \bibinfo {author} {\bibfnamefont {P.}~\bibnamefont
  {Gaspard}},\ }\bibfield  {title} {\bibinfo {title} {The fluctuation theorem
  for currents in semi-markov processes},\ }\href
  {https://doi.org/10.1088/1742-5468/2008/11/P11007} {\bibfield  {journal}
  {\bibinfo  {journal} {Journal of Statistical Mechanics: Theory and
  Experiment}\ }\textbf {\bibinfo {volume} {2008}},\ \bibinfo {pages} {P11007}
  (\bibinfo {year} {2008})}\BibitemShut {NoStop}%
\bibitem [{\citenamefont {Schnakenberg}(1976)}]{schnakenberg1976network}%
  \BibitemOpen
  \bibfield  {author} {\bibinfo {author} {\bibfnamefont {J.}~\bibnamefont
  {Schnakenberg}},\ }\bibfield  {title} {\bibinfo {title} {Network theory of
  microscopic and macroscopic behavior of master equation systems},\
  }\href@noop {} {\bibfield  {journal} {\bibinfo  {journal} {Reviews of Modern
  physics}\ }\textbf {\bibinfo {volume} {48}},\ \bibinfo {pages} {571}
  (\bibinfo {year} {1976})}\BibitemShut {NoStop}%
\bibitem [{\citenamefont {Van~den Broeck}\ and\ \citenamefont
  {Esposito}(2015)}]{van2015ensemble}%
  \BibitemOpen
  \bibfield  {author} {\bibinfo {author} {\bibfnamefont {C.}~\bibnamefont
  {Van~den Broeck}}\ and\ \bibinfo {author} {\bibfnamefont {M.}~\bibnamefont
  {Esposito}},\ }\bibfield  {title} {\bibinfo {title} {Ensemble and trajectory
  thermodynamics: A brief introduction},\ }\href@noop {} {\bibfield  {journal}
  {\bibinfo  {journal} {Physica A: Statistical Mechanics and its Applications}\
  }\textbf {\bibinfo {volume} {418}},\ \bibinfo {pages} {6} (\bibinfo {year}
  {2015})}\BibitemShut {NoStop}%
\bibitem [{\citenamefont {Mallick}(2009)}]{mallick2009some}%
  \BibitemOpen
  \bibfield  {author} {\bibinfo {author} {\bibfnamefont {K.}~\bibnamefont
  {Mallick}},\ }\bibfield  {title} {\bibinfo {title} {Some recent developments
  in non-equilibrium statistical physics},\ }\href@noop {} {\bibfield
  {journal} {\bibinfo  {journal} {Pramana}\ }\textbf {\bibinfo {volume} {73}},\
  \bibinfo {pages} {417} (\bibinfo {year} {2009})}\BibitemShut {NoStop}%
\bibitem [{\citenamefont {Peliti}\ and\ \citenamefont
  {Pigolotti}(2021)}]{peliti2021stochastic}%
  \BibitemOpen
  \bibfield  {author} {\bibinfo {author} {\bibfnamefont {L.}~\bibnamefont
  {Peliti}}\ and\ \bibinfo {author} {\bibfnamefont {S.}~\bibnamefont
  {Pigolotti}},\ }\href@noop {} {\emph {\bibinfo {title} {Stochastic
  Thermodynamics: An Introduction}}}\ (\bibinfo  {publisher} {Princeton
  University Press},\ \bibinfo {year} {2021})\BibitemShut {NoStop}%
\bibitem [{\citenamefont {Ziener}\ \emph {et~al.}(2015)\citenamefont {Ziener},
  \citenamefont {Maritan},\ and\ \citenamefont
  {Hinrichsen}}]{ziener2015entropy}%
  \BibitemOpen
  \bibfield  {author} {\bibinfo {author} {\bibfnamefont {R.}~\bibnamefont
  {Ziener}}, \bibinfo {author} {\bibfnamefont {A.}~\bibnamefont {Maritan}},\
  and\ \bibinfo {author} {\bibfnamefont {H.}~\bibnamefont {Hinrichsen}},\
  }\bibfield  {title} {\bibinfo {title} {On entropy production in
  nonequilibrium systems},\ }\href@noop {} {\bibfield  {journal} {\bibinfo
  {journal} {Journal of Statistical Mechanics: Theory and Experiment}\ }\textbf
  {\bibinfo {volume} {2015}},\ \bibinfo {pages} {P08014} (\bibinfo {year}
  {2015})}\BibitemShut {NoStop}%
\bibitem [{\citenamefont {Gaspard}(2004)}]{gaspard2004time}%
  \BibitemOpen
  \bibfield  {author} {\bibinfo {author} {\bibfnamefont {P.}~\bibnamefont
  {Gaspard}},\ }\bibfield  {title} {\bibinfo {title} {Time-reversed dynamical
  entropy and irreversibility in markovian random processes},\ }\href@noop {}
  {\bibfield  {journal} {\bibinfo  {journal} {Journal of statistical physics}\
  }\textbf {\bibinfo {volume} {117}},\ \bibinfo {pages} {599} (\bibinfo {year}
  {2004})}\BibitemShut {NoStop}%
\bibitem [{\citenamefont {Maes}\ and\ \citenamefont
  {Neto{\v{c}}n{\`y}}(2003)}]{maes2003time}%
  \BibitemOpen
  \bibfield  {author} {\bibinfo {author} {\bibfnamefont {C.}~\bibnamefont
  {Maes}}\ and\ \bibinfo {author} {\bibfnamefont {K.}~\bibnamefont
  {Neto{\v{c}}n{\`y}}},\ }\bibfield  {title} {\bibinfo {title} {Time-reversal
  and entropy},\ }\href@noop {} {\bibfield  {journal} {\bibinfo  {journal}
  {Journal of statistical physics}\ }\textbf {\bibinfo {volume} {110}},\
  \bibinfo {pages} {269} (\bibinfo {year} {2003})}\BibitemShut {NoStop}%
\bibitem [{\citenamefont {Parrondo}\ \emph {et~al.}(2009)\citenamefont
  {Parrondo}, \citenamefont {Van~den Broeck},\ and\ \citenamefont
  {Kawai}}]{parrondo2009entropy}%
  \BibitemOpen
  \bibfield  {author} {\bibinfo {author} {\bibfnamefont {J.~M.}\ \bibnamefont
  {Parrondo}}, \bibinfo {author} {\bibfnamefont {C.}~\bibnamefont {Van~den
  Broeck}},\ and\ \bibinfo {author} {\bibfnamefont {R.}~\bibnamefont {Kawai}},\
  }\bibfield  {title} {\bibinfo {title} {Entropy production and the arrow of
  time},\ }\href@noop {} {\bibfield  {journal} {\bibinfo  {journal} {New
  Journal of Physics}\ }\textbf {\bibinfo {volume} {11}},\ \bibinfo {pages}
  {073008} (\bibinfo {year} {2009})}\BibitemShut {NoStop}%
\bibitem [{\citenamefont {Varadhan}(2010)}]{varadhan2010large}%
  \BibitemOpen
  \bibfield  {author} {\bibinfo {author} {\bibfnamefont {S.~S.}\ \bibnamefont
  {Varadhan}},\ }\bibfield  {title} {\bibinfo {title} {Large deviations},\ }in\
  \href {https://doi.org/10.1142/9789814324359_0027} {\emph {\bibinfo
  {booktitle} {Proceedings of the International Congress of Mathematicians 2010
  (ICM 2010) (In 4 Volumes) Vol. I: Plenary Lectures and Ceremonies Vols.
  II--IV: Invited Lectures}}}\ (\bibinfo {organization} {World Scientific},\
  \bibinfo {year} {2010})\ pp.\ \bibinfo {pages} {622--639}\BibitemShut
  {NoStop}%
\bibitem [{\citenamefont {Touchette}(2011)}]{touchette2011basic}%
  \BibitemOpen
  \bibfield  {author} {\bibinfo {author} {\bibfnamefont {H.}~\bibnamefont
  {Touchette}},\ }\bibfield  {title} {\bibinfo {title} {A basic introduction to
  large deviations: Theory, applications, simulations},\ }\bibfield  {journal}
  {\bibinfo  {journal} {arXiv preprint arXiv:1106.4146}\ }\href
  {https://doi.org/10.48550/arXiv.1106.4146} {10.48550/arXiv.1106.4146}
  (\bibinfo {year} {2011})\BibitemShut {NoStop}%
\bibitem [{\citenamefont {Mehl}\ \emph {et~al.}(2012)\citenamefont {Mehl},
  \citenamefont {Lander}, \citenamefont {Bechinger}, \citenamefont {Blickle},\
  and\ \citenamefont {Seifert}}]{mehl2012role}%
  \BibitemOpen
  \bibfield  {author} {\bibinfo {author} {\bibfnamefont {J.}~\bibnamefont
  {Mehl}}, \bibinfo {author} {\bibfnamefont {B.}~\bibnamefont {Lander}},
  \bibinfo {author} {\bibfnamefont {C.}~\bibnamefont {Bechinger}}, \bibinfo
  {author} {\bibfnamefont {V.}~\bibnamefont {Blickle}},\ and\ \bibinfo {author}
  {\bibfnamefont {U.}~\bibnamefont {Seifert}},\ }\bibfield  {title} {\bibinfo
  {title} {Role of hidden slow degrees of freedom in the fluctuation theorem},\
  }\href {https://doi.org/10.1103/PhysRevLett.108.220601} {\bibfield  {journal}
  {\bibinfo  {journal} {Phys. Rev. Lett.}\ }\textbf {\bibinfo {volume} {108}},\
  \bibinfo {pages} {220601} (\bibinfo {year} {2012})}\BibitemShut {NoStop}%
\bibitem [{\citenamefont {Battle}\ \emph {et~al.}(2016)\citenamefont {Battle},
  \citenamefont {Broedersz}, \citenamefont {Fakhri}, \citenamefont {Geyer},
  \citenamefont {Howard}, \citenamefont {Schmidt},\ and\ \citenamefont
  {MacKintosh}}]{battle2016broken}%
  \BibitemOpen
  \bibfield  {author} {\bibinfo {author} {\bibfnamefont {C.}~\bibnamefont
  {Battle}}, \bibinfo {author} {\bibfnamefont {C.~P.}\ \bibnamefont
  {Broedersz}}, \bibinfo {author} {\bibfnamefont {N.}~\bibnamefont {Fakhri}},
  \bibinfo {author} {\bibfnamefont {V.~F.}\ \bibnamefont {Geyer}}, \bibinfo
  {author} {\bibfnamefont {J.}~\bibnamefont {Howard}}, \bibinfo {author}
  {\bibfnamefont {C.~F.}\ \bibnamefont {Schmidt}},\ and\ \bibinfo {author}
  {\bibfnamefont {F.~C.}\ \bibnamefont {MacKintosh}},\ }\bibfield  {title}
  {\bibinfo {title} {Broken detailed balance at mesoscopic scales in active
  biological systems},\ }\href {https://doi.org/10.1126/science.aac8167}
  {\bibfield  {journal} {\bibinfo  {journal} {Science}\ }\textbf {\bibinfo
  {volume} {352}},\ \bibinfo {pages} {604} (\bibinfo {year}
  {2016})}\BibitemShut {NoStop}%
\bibitem [{\citenamefont {Di~Terlizzi}\ \emph {et~al.}(2024)\citenamefont
  {Di~Terlizzi}, \citenamefont {Gironella}, \citenamefont {Herraez-Aguilar},
  \citenamefont {Betz}, \citenamefont {Monroy}, \citenamefont {Baiesi},\ and\
  \citenamefont {Ritort}}]{diterlizzi2024variance}%
  \BibitemOpen
  \bibfield  {author} {\bibinfo {author} {\bibfnamefont {I.}~\bibnamefont
  {Di~Terlizzi}}, \bibinfo {author} {\bibfnamefont {M.}~\bibnamefont
  {Gironella}}, \bibinfo {author} {\bibfnamefont {D.}~\bibnamefont
  {Herraez-Aguilar}}, \bibinfo {author} {\bibfnamefont {T.}~\bibnamefont
  {Betz}}, \bibinfo {author} {\bibfnamefont {F.}~\bibnamefont {Monroy}},
  \bibinfo {author} {\bibfnamefont {M.}~\bibnamefont {Baiesi}},\ and\ \bibinfo
  {author} {\bibfnamefont {F.}~\bibnamefont {Ritort}},\ }\bibfield  {title}
  {\bibinfo {title} {Variance sum rule for entropy production},\ }\href
  {https://doi.org/10.1126/science.adh1823} {\bibfield  {journal} {\bibinfo
  {journal} {Science}\ }\textbf {\bibinfo {volume} {383}},\ \bibinfo {pages}
  {971} (\bibinfo {year} {2024})}\BibitemShut {NoStop}%
\bibitem [{\citenamefont {Touchette}(2009)}]{touchette2009large}%
  \BibitemOpen
  \bibfield  {author} {\bibinfo {author} {\bibfnamefont {H.}~\bibnamefont
  {Touchette}},\ }\bibfield  {title} {\bibinfo {title} {The large deviation
  approach to statistical mechanics},\ }\href
  {https://doi.org/https://doi.org/10.1016/j.physrep.2009.05.002} {\bibfield
  {journal} {\bibinfo  {journal} {Physics Reports}\ }\textbf {\bibinfo {volume}
  {478}},\ \bibinfo {pages} {1} (\bibinfo {year} {2009})}\BibitemShut {NoStop}%
\bibitem [{\citenamefont {Maes}(2021)}]{maes2021local}%
  \BibitemOpen
  \bibfield  {author} {\bibinfo {author} {\bibfnamefont {C.}~\bibnamefont
  {Maes}},\ }\bibfield  {title} {\bibinfo {title} {{Local detailed balance}},\
  }\href {https://doi.org/10.21468/SciPostPhysLectNotes.32} {\bibfield
  {journal} {\bibinfo  {journal} {SciPost Phys. Lect. Notes}\ ,\ \bibinfo
  {pages} {32}} (\bibinfo {year} {2021})}\BibitemShut {NoStop}%
\bibitem [{\citenamefont {Teza}\ \emph {et~al.}(2020)\citenamefont {Teza},
  \citenamefont {Iubini}, \citenamefont {Baiesi}, \citenamefont {Stella},\ and\
  \citenamefont {Vanderzande}}]{teza2020rate}%
  \BibitemOpen
  \bibfield  {author} {\bibinfo {author} {\bibfnamefont {G.}~\bibnamefont
  {Teza}}, \bibinfo {author} {\bibfnamefont {S.}~\bibnamefont {Iubini}},
  \bibinfo {author} {\bibfnamefont {M.}~\bibnamefont {Baiesi}}, \bibinfo
  {author} {\bibfnamefont {A.~L.}\ \bibnamefont {Stella}},\ and\ \bibinfo
  {author} {\bibfnamefont {C.}~\bibnamefont {Vanderzande}},\ }\bibfield
  {title} {\bibinfo {title} {Rate dependence of current and fluctuations in
  jump models with negative differential mobility},\ }\href
  {https://doi.org/https://doi.org/10.1016/j.physa.2019.123176} {\bibfield
  {journal} {\bibinfo  {journal} {Physica A: Statistical Mechanics and its
  Applications}\ }\textbf {\bibinfo {volume} {552}},\ \bibinfo {pages} {123176}
  (\bibinfo {year} {2020})},\ \bibinfo {note} {tributes of Non-equilibrium
  Statistical Physics}\BibitemShut {NoStop}%
\bibitem [{\citenamefont {Stella}\ \emph
  {et~al.}(2023{\natexlab{a}})\citenamefont {Stella}, \citenamefont
  {Chechkin},\ and\ \citenamefont {Teza}}]{stella2023anomalous}%
  \BibitemOpen
  \bibfield  {author} {\bibinfo {author} {\bibfnamefont {A.~L.}\ \bibnamefont
  {Stella}}, \bibinfo {author} {\bibfnamefont {A.}~\bibnamefont {Chechkin}},\
  and\ \bibinfo {author} {\bibfnamefont {G.}~\bibnamefont {Teza}},\ }\bibfield
  {title} {\bibinfo {title} {Anomalous dynamical scaling determines universal
  critical singularities},\ }\href
  {https://doi.org/10.1103/PhysRevLett.130.207104} {\bibfield  {journal}
  {\bibinfo  {journal} {Phys. Rev. Lett.}\ }\textbf {\bibinfo {volume} {130}},\
  \bibinfo {pages} {207104} (\bibinfo {year} {2023}{\natexlab{a}})}\BibitemShut
  {NoStop}%
\bibitem [{\citenamefont {Stella}\ \emph
  {et~al.}(2023{\natexlab{b}})\citenamefont {Stella}, \citenamefont
  {Chechkin},\ and\ \citenamefont {Teza}}]{stella2023universal}%
  \BibitemOpen
  \bibfield  {author} {\bibinfo {author} {\bibfnamefont {A.~L.}\ \bibnamefont
  {Stella}}, \bibinfo {author} {\bibfnamefont {A.}~\bibnamefont {Chechkin}},\
  and\ \bibinfo {author} {\bibfnamefont {G.}~\bibnamefont {Teza}},\ }\bibfield
  {title} {\bibinfo {title} {Universal singularities of anomalous diffusion in
  the richardson class},\ }\href {https://doi.org/10.1103/PhysRevE.107.054118}
  {\bibfield  {journal} {\bibinfo  {journal} {Phys. Rev. E}\ }\textbf {\bibinfo
  {volume} {107}},\ \bibinfo {pages} {054118} (\bibinfo {year}
  {2023}{\natexlab{b}})}\BibitemShut {NoStop}%
\bibitem [{\citenamefont {Teza}\ and\ \citenamefont
  {Stella}(2025)}]{teza2025universal}%
  \BibitemOpen
  \bibfield  {author} {\bibinfo {author} {\bibfnamefont {G.}~\bibnamefont
  {Teza}}\ and\ \bibinfo {author} {\bibfnamefont {A.~L.}\ \bibnamefont
  {Stella}},\ }\bibfield  {title} {\bibinfo {title} {Universal and nonuniversal
  signatures in the scaling functions of critical variables},\ }\href
  {https://doi.org/10.1103/PhysRevLett.134.127102} {\bibfield  {journal}
  {\bibinfo  {journal} {Phys. Rev. Lett.}\ }\textbf {\bibinfo {volume} {134}},\
  \bibinfo {pages} {127102} (\bibinfo {year} {2025})}\BibitemShut {NoStop}%
\bibitem [{SM()}]{SM}%
  \BibitemOpen
  \href@noop {} {\bibinfo {title} {See supplemental material at ... for
  additional details of the calculations at the basis of the results presented
  in the main text.}}\BibitemShut {Stop}%
\bibitem [{\citenamefont {Montroll}\ and\ \citenamefont
  {Weiss}(1965)}]{montroll1965random}%
  \BibitemOpen
  \bibfield  {author} {\bibinfo {author} {\bibfnamefont {E.~W.}\ \bibnamefont
  {Montroll}}\ and\ \bibinfo {author} {\bibfnamefont {G.~H.}\ \bibnamefont
  {Weiss}},\ }\bibfield  {title} {\bibinfo {title} {Random walks on lattices.
  ii},\ }\href {https://doi.org/10.1063/1.1704269} {\bibfield  {journal}
  {\bibinfo  {journal} {Journal of Mathematical Physics}\ }\textbf {\bibinfo
  {volume} {6}},\ \bibinfo {pages} {167} (\bibinfo {year} {1965})}\BibitemShut
  {NoStop}%
\bibitem [{\citenamefont {Klafter}\ and\ \citenamefont
  {Silbey}(1980)}]{klafter1980derivation}%
  \BibitemOpen
  \bibfield  {author} {\bibinfo {author} {\bibfnamefont {J.}~\bibnamefont
  {Klafter}}\ and\ \bibinfo {author} {\bibfnamefont {R.}~\bibnamefont
  {Silbey}},\ }\bibfield  {title} {\bibinfo {title} {Derivation of the
  continuous-time random-walk equation},\ }\href
  {https://doi.org/10.1103/PhysRevLett.44.55} {\bibfield  {journal} {\bibinfo
  {journal} {Phys. Rev. Lett.}\ }\textbf {\bibinfo {volume} {44}},\ \bibinfo
  {pages} {55} (\bibinfo {year} {1980})}\BibitemShut {NoStop}%
\bibitem [{\citenamefont {Di~Terlizzi}\ \emph {et~al.}(2020)\citenamefont
  {Di~Terlizzi}, \citenamefont {Ritort},\ and\ \citenamefont
  {Baiesi}}]{diterlizzi2020explicit}%
  \BibitemOpen
  \bibfield  {author} {\bibinfo {author} {\bibfnamefont {I.}~\bibnamefont
  {Di~Terlizzi}}, \bibinfo {author} {\bibfnamefont {F.}~\bibnamefont
  {Ritort}},\ and\ \bibinfo {author} {\bibfnamefont {M.}~\bibnamefont
  {Baiesi}},\ }\bibfield  {title} {\bibinfo {title} {Explicit solution of the
  generalised langevin equation},\ }\href
  {https://doi.org/10.1007/s10955-020-02639-4} {\bibfield  {journal} {\bibinfo
  {journal} {Journal of Statistical Physics}\ }\textbf {\bibinfo {volume}
  {181}},\ \bibinfo {pages} {1609} (\bibinfo {year} {2020})}\BibitemShut
  {NoStop}%
\bibitem [{\citenamefont {Dembo}(2009)}]{dembo2009large}%
  \BibitemOpen
  \bibfield  {author} {\bibinfo {author} {\bibfnamefont {A.}~\bibnamefont
  {Dembo}},\ }\href@noop {} {\emph {\bibinfo {title} {Large deviations
  techniques and applications}}}\ (\bibinfo  {publisher} {Springer},\ \bibinfo
  {year} {2009})\BibitemShut {NoStop}%
\bibitem [{\citenamefont {Greven}\ and\ \citenamefont {den
  Hollander}(1994)}]{greven1994large}%
  \BibitemOpen
  \bibfield  {author} {\bibinfo {author} {\bibfnamefont {A.}~\bibnamefont
  {Greven}}\ and\ \bibinfo {author} {\bibfnamefont {F.}~\bibnamefont {den
  Hollander}},\ }\bibfield  {title} {\bibinfo {title} {Large deviations for a
  random walk in random environment},\ }\href@noop {} {\bibfield  {journal}
  {\bibinfo  {journal} {The Annals of Probability}\ ,\ \bibinfo {pages} {1381}}
  (\bibinfo {year} {1994})}\BibitemShut {NoStop}%
\bibitem [{\citenamefont {Teza}(2020)}]{teza2020thesis}%
  \BibitemOpen
  \bibfield  {author} {\bibinfo {author} {\bibfnamefont {G.}~\bibnamefont
  {Teza}},\ }\emph {\bibinfo {title} {Out of equilibrium dynamics: from an
  entropy of the growth to the growth of entropy production}},\ \href@noop {}
  {Ph.D. thesis} (\bibinfo {year} {2020})\BibitemShut {NoStop}%
\bibitem [{\citenamefont {GrandPre}\ \emph {et~al.}(2024)\citenamefont
  {GrandPre}, \citenamefont {Teza},\ and\ \citenamefont
  {Bialek}}]{grandpre2024direct}%
  \BibitemOpen
  \bibfield  {author} {\bibinfo {author} {\bibfnamefont {T.}~\bibnamefont
  {GrandPre}}, \bibinfo {author} {\bibfnamefont {G.}~\bibnamefont {Teza}},\
  and\ \bibinfo {author} {\bibfnamefont {W.}~\bibnamefont {Bialek}},\
  }\bibfield  {title} {\bibinfo {title} {Direct estimates of irreversibility
  from time series},\ }\bibfield  {journal} {\bibinfo  {journal} {arXiv
  preprint arXiv:2412.19772}\ }\href
  {https://doi.org/10.48550/arXiv.2412.19772} {10.48550/arXiv.2412.19772}
  (\bibinfo {year} {2024})\BibitemShut {NoStop}%
\bibitem [{\citenamefont {G\"{a}rtner}(1977)}]{gartner1977on}%
  \BibitemOpen
  \bibfield  {author} {\bibinfo {author} {\bibfnamefont {J.}~\bibnamefont
  {G\"{a}rtner}},\ }\bibfield  {title} {\bibinfo {title} {On large deviations
  from the invariant measure},\ }\href {https://doi.org/10.1137/1122003}
  {\bibfield  {journal} {\bibinfo  {journal} {Theory of Probability \& Its
  Applications}\ }\textbf {\bibinfo {volume} {22}},\ \bibinfo {pages} {24}
  (\bibinfo {year} {1977})},\ \Eprint
  {https://arxiv.org/abs/https://doi.org/10.1137/1122003}
  {https://doi.org/10.1137/1122003} \BibitemShut {NoStop}%
\bibitem [{\citenamefont {Ellis}(1984)}]{ellis1984large}%
  \BibitemOpen
  \bibfield  {author} {\bibinfo {author} {\bibfnamefont {R.~S.}\ \bibnamefont
  {Ellis}},\ }\bibfield  {title} {\bibinfo {title} {Large deviations for a
  general class of random vectors},\ }\href
  {http://www.jstor.org/stable/2243592} {\bibfield  {journal} {\bibinfo
  {journal} {The Annals of Probability}\ }\textbf {\bibinfo {volume} {12}},\
  \bibinfo {pages} {1} (\bibinfo {year} {1984})}\BibitemShut {NoStop}%
\bibitem [{\citenamefont {Puglisi}\ \emph {et~al.}(2010)\citenamefont
  {Puglisi}, \citenamefont {Pigolotti}, \citenamefont {Rondoni},\ and\
  \citenamefont {Vulpiani}}]{puglisi2010entropy}%
  \BibitemOpen
  \bibfield  {author} {\bibinfo {author} {\bibfnamefont {A.}~\bibnamefont
  {Puglisi}}, \bibinfo {author} {\bibfnamefont {S.}~\bibnamefont {Pigolotti}},
  \bibinfo {author} {\bibfnamefont {L.}~\bibnamefont {Rondoni}},\ and\ \bibinfo
  {author} {\bibfnamefont {A.}~\bibnamefont {Vulpiani}},\ }\bibfield  {title}
  {\bibinfo {title} {Entropy production and coarse graining in markov
  processes},\ }\href {https://doi.org/10.1088/1742-5468/2010/05/P05015}
  {\bibfield  {journal} {\bibinfo  {journal} {Journal of Statistical Mechanics:
  Theory and Experiment}\ }\textbf {\bibinfo {volume} {2010}},\ \bibinfo
  {pages} {P05015} (\bibinfo {year} {2010})}\BibitemShut {NoStop}%
\bibitem [{\citenamefont {Bo}\ and\ \citenamefont
  {Celani}(2014)}]{bo2014entropy}%
  \BibitemOpen
  \bibfield  {author} {\bibinfo {author} {\bibfnamefont {S.}~\bibnamefont
  {Bo}}\ and\ \bibinfo {author} {\bibfnamefont {A.}~\bibnamefont {Celani}},\
  }\bibfield  {title} {\bibinfo {title} {Entropy production in stochastic
  systems with fast and slow time-scales},\ }\href
  {https://doi.org/10.1007/s10955-014-0922-1} {\bibfield  {journal} {\bibinfo
  {journal} {Journal of Statistical Physics}\ }\textbf {\bibinfo {volume}
  {154}},\ \bibinfo {pages} {1325} (\bibinfo {year} {2014})}\BibitemShut
  {NoStop}%
\bibitem [{\citenamefont {Van~Vu}\ and\ \citenamefont
  {Hasegawa}(2019)}]{van2019uncertainty}%
  \BibitemOpen
  \bibfield  {author} {\bibinfo {author} {\bibfnamefont {T.}~\bibnamefont
  {Van~Vu}}\ and\ \bibinfo {author} {\bibfnamefont {Y.}~\bibnamefont
  {Hasegawa}},\ }\bibfield  {title} {\bibinfo {title} {Uncertainty relations
  for underdamped langevin dynamics},\ }\href@noop {} {\bibfield  {journal}
  {\bibinfo  {journal} {Physical Review E}\ }\textbf {\bibinfo {volume}
  {100}},\ \bibinfo {pages} {032130} (\bibinfo {year} {2019})}\BibitemShut
  {NoStop}%
\bibitem [{\citenamefont {Fischer}\ \emph {et~al.}(2018)\citenamefont
  {Fischer}, \citenamefont {Pietzonka},\ and\ \citenamefont
  {Seifert}}]{fischer2018large}%
  \BibitemOpen
  \bibfield  {author} {\bibinfo {author} {\bibfnamefont {L.~P.}\ \bibnamefont
  {Fischer}}, \bibinfo {author} {\bibfnamefont {P.}~\bibnamefont {Pietzonka}},\
  and\ \bibinfo {author} {\bibfnamefont {U.}~\bibnamefont {Seifert}},\
  }\bibfield  {title} {\bibinfo {title} {Large deviation function for a driven
  underdamped particle in a periodic potential},\ }\href@noop {} {\bibfield
  {journal} {\bibinfo  {journal} {Physical Review E}\ }\textbf {\bibinfo
  {volume} {97}},\ \bibinfo {pages} {022143} (\bibinfo {year}
  {2018})}\BibitemShut {NoStop}%
\bibitem [{\citenamefont {Pietzonka}(2022)}]{pietzonka2022classical}%
  \BibitemOpen
  \bibfield  {author} {\bibinfo {author} {\bibfnamefont {P.}~\bibnamefont
  {Pietzonka}},\ }\bibfield  {title} {\bibinfo {title} {Classical pendulum
  clocks break the thermodynamic uncertainty relation},\ }\href@noop {}
  {\bibfield  {journal} {\bibinfo  {journal} {Physical Review Letters}\
  }\textbf {\bibinfo {volume} {128}},\ \bibinfo {pages} {130606} (\bibinfo
  {year} {2022})}\BibitemShut {NoStop}%
\bibitem [{\citenamefont {Koyuk}\ and\ \citenamefont
  {Seifert}(2020)}]{koyuk2020thermodynamic}%
  \BibitemOpen
  \bibfield  {author} {\bibinfo {author} {\bibfnamefont {T.}~\bibnamefont
  {Koyuk}}\ and\ \bibinfo {author} {\bibfnamefont {U.}~\bibnamefont
  {Seifert}},\ }\bibfield  {title} {\bibinfo {title} {Thermodynamic uncertainty
  relation for time-dependent driving},\ }\href@noop {} {\bibfield  {journal}
  {\bibinfo  {journal} {Physical Review Letters}\ }\textbf {\bibinfo {volume}
  {125}},\ \bibinfo {pages} {260604} (\bibinfo {year} {2020})}\BibitemShut
  {NoStop}%
\bibitem [{\citenamefont {Macieszczak}\ \emph {et~al.}(2018)\citenamefont
  {Macieszczak}, \citenamefont {Brandner},\ and\ \citenamefont
  {Garrahan}}]{macieszczak2018unified}%
  \BibitemOpen
  \bibfield  {author} {\bibinfo {author} {\bibfnamefont {K.}~\bibnamefont
  {Macieszczak}}, \bibinfo {author} {\bibfnamefont {K.}~\bibnamefont
  {Brandner}},\ and\ \bibinfo {author} {\bibfnamefont {J.~P.}\ \bibnamefont
  {Garrahan}},\ }\bibfield  {title} {\bibinfo {title} {Unified thermodynamic
  uncertainty relations in linear response},\ }\href@noop {} {\bibfield
  {journal} {\bibinfo  {journal} {Physical review letters}\ }\textbf {\bibinfo
  {volume} {121}},\ \bibinfo {pages} {130601} (\bibinfo {year}
  {2018})}\BibitemShut {NoStop}%
\bibitem [{\citenamefont {Dieball}\ and\ \citenamefont
  {Godec}(2023)}]{dieball2023direct}%
  \BibitemOpen
  \bibfield  {author} {\bibinfo {author} {\bibfnamefont {C.}~\bibnamefont
  {Dieball}}\ and\ \bibinfo {author} {\bibfnamefont {A.}~\bibnamefont
  {Godec}},\ }\bibfield  {title} {\bibinfo {title} {Direct route to
  thermodynamic uncertainty relations and their saturation},\ }\href@noop {}
  {\bibfield  {journal} {\bibinfo  {journal} {Physical Review Letters}\
  }\textbf {\bibinfo {volume} {130}},\ \bibinfo {pages} {087101} (\bibinfo
  {year} {2023})}\BibitemShut {NoStop}%
\bibitem [{\citenamefont {Liu}\ \emph {et~al.}(2020)\citenamefont {Liu},
  \citenamefont {Gong},\ and\ \citenamefont {Ueda}}]{liu2020thermodynamic}%
  \BibitemOpen
  \bibfield  {author} {\bibinfo {author} {\bibfnamefont {K.}~\bibnamefont
  {Liu}}, \bibinfo {author} {\bibfnamefont {Z.}~\bibnamefont {Gong}},\ and\
  \bibinfo {author} {\bibfnamefont {M.}~\bibnamefont {Ueda}},\ }\bibfield
  {title} {\bibinfo {title} {Thermodynamic uncertainty relation for arbitrary
  initial states},\ }\href@noop {} {\bibfield  {journal} {\bibinfo  {journal}
  {Physical Review Letters}\ }\textbf {\bibinfo {volume} {125}},\ \bibinfo
  {pages} {140602} (\bibinfo {year} {2020})}\BibitemShut {NoStop}%
\bibitem [{\citenamefont {Dechant}\ and\ \citenamefont
  {Sasa}(2018)}]{dechant2018current}%
  \BibitemOpen
  \bibfield  {author} {\bibinfo {author} {\bibfnamefont {A.}~\bibnamefont
  {Dechant}}\ and\ \bibinfo {author} {\bibfnamefont {S.-i.}\ \bibnamefont
  {Sasa}},\ }\bibfield  {title} {\bibinfo {title} {Current fluctuations and
  transport efficiency for general langevin systems},\ }\href@noop {}
  {\bibfield  {journal} {\bibinfo  {journal} {Journal of Statistical Mechanics:
  Theory and Experiment}\ }\textbf {\bibinfo {volume} {2018}},\ \bibinfo
  {pages} {063209} (\bibinfo {year} {2018})}\BibitemShut {NoStop}%
\bibitem [{\citenamefont {Koyuk}\ and\ \citenamefont
  {Seifert}(2019)}]{koyuk2019operationally}%
  \BibitemOpen
  \bibfield  {author} {\bibinfo {author} {\bibfnamefont {T.}~\bibnamefont
  {Koyuk}}\ and\ \bibinfo {author} {\bibfnamefont {U.}~\bibnamefont
  {Seifert}},\ }\bibfield  {title} {\bibinfo {title} {Operationally accessible
  bounds on fluctuations and entropy production in periodically driven
  systems},\ }\href@noop {} {\bibfield  {journal} {\bibinfo  {journal}
  {Physical review letters}\ }\textbf {\bibinfo {volume} {122}},\ \bibinfo
  {pages} {230601} (\bibinfo {year} {2019})}\BibitemShut {NoStop}%
\bibitem [{\citenamefont {Proesmans}\ and\ \citenamefont {Van~den
  Broeck}(2017)}]{proesmans2017discrete}%
  \BibitemOpen
  \bibfield  {author} {\bibinfo {author} {\bibfnamefont {K.}~\bibnamefont
  {Proesmans}}\ and\ \bibinfo {author} {\bibfnamefont {C.}~\bibnamefont
  {Van~den Broeck}},\ }\bibfield  {title} {\bibinfo {title} {Discrete-time
  thermodynamic uncertainty relation},\ }\href@noop {} {\bibfield  {journal}
  {\bibinfo  {journal} {Europhysics Letters}\ }\textbf {\bibinfo {volume}
  {119}},\ \bibinfo {pages} {20001} (\bibinfo {year} {2017})}\BibitemShut
  {NoStop}%
\bibitem [{\citenamefont {GrandPre}\ \emph {et~al.}(2021)\citenamefont
  {GrandPre}, \citenamefont {Klymko}, \citenamefont {Mandadapu},\ and\
  \citenamefont {Limmer}}]{grandpre2021entropy}%
  \BibitemOpen
  \bibfield  {author} {\bibinfo {author} {\bibfnamefont {T.}~\bibnamefont
  {GrandPre}}, \bibinfo {author} {\bibfnamefont {K.}~\bibnamefont {Klymko}},
  \bibinfo {author} {\bibfnamefont {K.~K.}\ \bibnamefont {Mandadapu}},\ and\
  \bibinfo {author} {\bibfnamefont {D.~T.}\ \bibnamefont {Limmer}},\ }\bibfield
   {title} {\bibinfo {title} {Entropy production fluctuations encode collective
  behavior in active matter},\ }\href@noop {} {\bibfield  {journal} {\bibinfo
  {journal} {Physical Review E}\ }\textbf {\bibinfo {volume} {103}},\ \bibinfo
  {pages} {012613} (\bibinfo {year} {2021})}\BibitemShut {NoStop}%
\bibitem [{\citenamefont {Wu}\ and\ \citenamefont
  {Jia}(2025)}]{wu2025parameter}%
  \BibitemOpen
  \bibfield  {author} {\bibinfo {author} {\bibfnamefont {B.}~\bibnamefont
  {Wu}}\ and\ \bibinfo {author} {\bibfnamefont {C.}~\bibnamefont {Jia}},\
  }\bibfield  {title} {\bibinfo {title} {Parameter inference and nonequilibrium
  identification for markov networks based on coarse-grained observations},\
  }\href@noop {} {\bibfield  {journal} {\bibinfo  {journal} {Physical Review
  Letters}\ }\textbf {\bibinfo {volume} {134}},\ \bibinfo {pages} {087103}
  (\bibinfo {year} {2025})}\BibitemShut {NoStop}%
\end{thebibliography}
\end{document}